\documentclass[twocolumn,aps,prd,preprintnumbers,showpacs,superscriptaddress,nofootinbib,amsmath,amssymb,floats,floatfix,showkeys,notitlepage]{revtex4}

\usepackage{graphicx}
\usepackage{subfigure}
\usepackage{palatino}
\usepackage{changes}
\usepackage{hyperref}
\hypersetup{colorlinks=true,linkcolor=blue,urlcolor=blue,citecolor=blue}
\usepackage[toc,page]{appendix}
\usepackage[normalem]{ulem}
\usepackage{adjustbox}
\usepackage{latexsym}
\usepackage{amsmath}
\usepackage{amssymb}
\usepackage{amsfonts}
\usepackage{times}
\usepackage{dcolumn}
\usepackage{bm}
\usepackage{tikz}
\usepackage{bigints}
\usepackage{array,tabularx,multirow,booktabs}
\usepackage{tabularx}
\usepackage{multirow}
\usepackage[tracking=true]{microtype}
\SetTracking{}{500}
\SetTracking{encoding={*}, shape=sc}{40}
\allowdisplaybreaks

\begin{document} \sloppy
\title{Shadow and weak deflection angle of extended uncertainty principle black hole \protect \\surrounded with dark matter}
\author{Reggie C. Pantig}
\email{reggie.pantig@dlsu.edu.ph}
\affiliation{Physics Department, De La Salle University, 2401 Taft Avenue, Manila, 1004 Philippines}
\author{Paul K. Yu}
\email{paul\_k\_yu@dlsu.edu.ph}
\affiliation{Physics Department, De La Salle University, 2401 Taft Avenue, Manila, 1004 Philippines}
\author{Emmanuel T. Rodulfo}
\email{emmanuel.rodulfo@dlsu.edu.ph}
\affiliation{Physics Department, De La Salle University, 2401 Taft Avenue, Manila, 1004 Philippines}
\author{Ali \"Ovg\"un}
\email{ali.ovgun@emu.edu.tr}
\homepage{https://www.aovgun.com}
\affiliation{Physics Department, Eastern Mediterranean University, Famagusta, 99628 North Cyprus via Mersin 10, Turkey}

\begin{abstract}
In this paper, we discuss the possible effects of dark matter on a Schwarzschild black hole with the correction of extended uncertainty principle (EUP), such as the parameter $\alpha$ and the large fundamental length scale $L_*$. In particular, we surround the EUP black hole of mass $m$ with a static spherical shell of dark matter described by the parameters mass $M$, inner radius $r_s$, and thickness $\Delta r_s$. In this study, we find that there is no deviation in the event horizon, which readily implies that the black hole temperature due to the Hawking radiation is independent of any dark matter concentration. In addition, we show some effects of the EUP parameter on the innermost stable circular orbit (ISCO) radius of time-like particles, photon sphere, shadow radius, and weak deflection angle. It is found that time-like orbits are affected by deviation of low values of mass $M$. A greater dark matter density is needed to have remarkable effects on the null orbits. Using the analytic expression for the shadow radius and the approximation $\Delta r_s>>r_s$, it is revealed that $L_*$ should not be lower than $2m$. To broaden the scope of this study, we also calculate the analytic expression for the weak deflection angle using the Ishihara et al. method \cite{Ishihara_2016}. As a result, we show that $\Delta r_s$ is improved by a factor of $(1+4\alpha m^2/L_*^2)$ due to the EUP correction parameters. The calculated shadow radius and weak deflection angle are then compared using the estimated values of the galactic mass from Sgr A*, M87, and UGC 7232, as well as the mass of the supermassive black hole at their center.

\end{abstract}



\maketitle

\section{Introduction} \label{sec1}
In 1915, Albert Einstein had formulated the general theory of relativity, which explains gravity within the geometry of spacetime \cite{Einstein_1916}. Later Karl Schwarzschild had found an exact solution of the Einstein field equations for static spherically symmetric compact object \cite{Schwarzschild_1916}, which is known as Schwarzschild black hole. 

Various researches have studied the shadow of the black holes, such as the shadow of the Schwarzschild black holes \cite{Synge_1966}, shadow of the Kerr black holes \cite{Bardeen_1973, Wei_2019}, shadow of the Kerr-like black holes \cite{Atamurotov_2014}, shadow of the Kaluza Klein black holes \cite{Nodehi_2020}, shadow of the naked singularities \cite{Bambhaniya_2021}, shadow of the Reissner-Nordstrom-de Sitter black holes \cite{Cotaescu_2021}, shadow of the Weyl black holes \cite{Fathi_2020}, and much more \cite{Atamurotov2016,Atamurotov2015,Papnoi2014,Atamurotov2013a,Atamurotov2013b,Abdujabbarov2013}. Black holes have been piquing so much interest among researchers since the Event Horizon Telescope (EHT) captured the first image of the shadow of the SMBH at the center of the M87 galaxy in 2019 \cite{Collaboration_2019}. Recently, the second image was captured which shows how an astrophysical environment, like the magnetic field, distorts the image of the black hole \cite{Event2021a,Event2021b}. It is further confirmed that, also within our galaxy, there is a supermassive black hole lurking and such work is pioneered by Ghez and Genzel \cite{Ghez_1998, Eckart_1997}. Mathematically, Roger Penrose discovered that black hole formation is a robust prediction of the general theory of relativity which made him, along with Ghez, awarded the 2020 Nobel prize in Physics \cite{Nobel_2020, Penrose_1965}.

The most mysterious phenomenon in modern cosmology – is the existence of dark matter. The $ \Lambda $CDM model of cosmology suggests that the content of our universe is made up of 27\% dark matter, which constitutes 85\% of the total mass \cite{Jarosik_2011} of the Universe. In an attempt to understand the effect or the possibility of detecting dark matter through the black hole geometry, several theoretical studies have been made \cite{Nikitin_2017,Xu_2018,Xu_2020}. Hou et al. studied the shadow of Sagittarius A* supermassive black hole in dark matter halo \cite{Hou_2018}; \"Ovg\"un studied weak deflection angle of photons through dark matter by black holes and wormholes using Gauss-Bonnet theorem \cite{Ovgun:2020yuv,Ovgun_2019}, whereas, others extended the studies of Hou et al. on the shadow of M87 supermassive black hole in dark matter halo \cite{Jusufi_2019}. Moreover, Haroon et al. and Hou et al. independently used a perfect fluid dark matter model to study rotating black hole shadows \cite{Haroon_2019,Hou_2018_2}. Konoplya sought a less model-dependent view on dark matter \cite{Konoplya_2019} and studied its effect on the shadow radius. Later on, Pantig et al. \cite{Pantig_2020_weak, Pantig_2020_rotating} extended Konoplya's analysis in determining the weak deflection angle, and applied the dark matter model to a rotating black hole.

There are several interests in the study of gravitational lensing to probe the existence of dark matter \cite{Trimble_1987,Kaiser_1993,Metcalf_2001,Virbhadra_2000,Kogan_2019}. The most popular method of calculating deflection angle by black holes, especially in asymptotically flat spacetime, is the Gauss-Bonnet theorem (GBT), pioneered by Gibbons and Werner \cite{Gibbons_2008,Gibbons_2009}. Since then, various studies have been explored the deflection angle by different black hole spacetimes, \cite{Ovgun_2018,Ovgun_2018_2,Ovgun_2018_3,Ovgun_2019,Ovgun_2019_2,Jusufi:2017mav,Ovgun_2019_4,Ovgun_2019_5,Jusufi:2017lsl,Sakalli:2017ewb,Jusufi:2017vta,Kumaran:2019qqp,Jusufi:2018jof,Javed:2019kon,Jusufi:2017hed,Li:2020dln,Jusufi:2017uhh,Javed:2019qyg,Jusufi:2018kmk,Lu2019,Li2019,Sereno2004,Babar2020,Atamurotov2021}. On the other hand, Ishihara et al. extended GBT to non-asymptotically flat spacetimes using the finite distance method \cite{Ishihara_2017,Ono_2017,Ono_2018,Ono_2019,Takizawa_2020,Li_2020,Li_2020_2,Pantig_2020_weak,Pantig_2020_rotating}.

Physicists have long used black holes as a way to probe the edges of the quantum nature of gravity \cite{Rovelli_2004}. Maggiore found a generalized uncertainty principle (GUP) using a Gedanken experiment of a black hole in a quantum gravity \cite{Maggiore_1993_1, Maggiore_1993_2}. Bambi et al. extended GUP to EUP \cite{Bambi_2008}, then the EUP effects are analyzed by various works \cite{Zhu_2009,Mignemi_2010,Costafilho_2016}. The EUP black hole is further developed by Mureika \cite{Mureika_2019}. The effect of the plasma medium on the deflection angle by EUP black hole was further analyzed in \cite{Kumaran:2019qqp}. In the quest of probing the appropriate large scale fundamental length, \cite{Lu_2019} studied the gravitational lensings on EUP black hole. Moreover, the temperature and Unruh effect of the EUP black hole was also explored in Ref. \cite{CHUNG2019451}.

In this paper, the main aim is to investigate the changes in the black hole geometry due to the dark matter mass and the EUP parameters. These deviations in the black hole geometry can affect the behavior of the null and time-like geodesics in the vicinity of the black hole so that it is interesting to study shadow radius and weak deflection angle. Hence, the goal is to estimate analytically dark matter thickness and see what the EUP parameters may bring to its fulfillment in our galaxy and others. In this study, we will use Konoplya's mass function \cite{Konoplya_2019} since it only advocates the basic properties of dark matter, thus, making it a less model-dependent one.

The paper is structured as follows: In Sect. \ref{sec2}, we briefly review the EUP black hole and introduce the mass function which contains information about the dark matter distribution. Further, the basic properties of the metric will be discussed. In Sect. \ref{sec3}, we derive an expression for the photon sphere and shadow radius and analyze it numerically. Sect. \ref{sec4} involves the analysis of the weak deflection angle using the finite distance approach by Ishihara  \cite{Ishihara_2016}. Finally, conclusive remarks and recommendations for future research directions are given in Sect. \ref{sec5}. In what follows, we used $G=c=1$, and the metric signature is $(-,+,+,+)$.

\section{Metric of EUP black hole surrounded by dark matter} \label{sec2}
Lets consider the black hole model recently proposed by Mureika \cite{Mureika_2019}, which incorporates the extended uncertainty principle that introduces a position-momentum uncertainty correction \cite{Bambi_2008}:
\begin{equation} \label{e1}
    \Delta x \Delta p \geq 1+\alpha \frac{\Delta x^2}{L_{*}^2},
\end{equation}
where a new large fundamental distance scale $ L_{*} $ is introduced, and the EUP parameter $ \alpha $ is usually taken in unity. The EUP provides quantum effects over macroscopic distances, suggesting that it may contribute to dark matter effects \cite{Mureika_2019}.

If the Schwarzschild event horizon is considered as the confinement of $N$ gravitons, each having momentum uncertainty $\Delta p_g$, then $\Delta x \sim 2m $ and we have
\begin{equation} \label{e2}
    \Delta p_{g} \sim \frac{\hbar }{2m} \left ( 1 + \frac{4\alpha m^{2}}{L_{*}^{2}} \right ).
\end{equation}
Here, $m$ is the mass of the black hole. Since $\frac{\hbar }{2m}$ represents the mass of each graviton, then $ N \frac{\hbar }{2m} \sim m $. Thus, the total momentum uncertainty is
\begin{equation} \label{e3}
    \Delta P \sim m \left ( 1 + \frac{4\alpha m^{2}}{L_{*}^{2}} \right ).
\end{equation}
Assuming that the stress-energy tensor has some EUP-correction \cite{Mureika_2019}, the ADM mass is given by
\begin{equation} \label{e4}
    m_{\text{EUP}} = \int d^{3}x \sqrt{g} \left (T_{0 \text{GR}}^{0} + T_{0 \text{EUP}}^{0} \right ),
\end{equation}
which is also $\Delta P$. Therefore,
\begin{equation} \label{e5}
    m_{\text{EUP}} = m \left ( 1 + \frac{4\alpha m^{2}}{L_{*}^{2}} \right ),
\end{equation}
which allows one to formulate the metric function of EUP-inspired Schwarzschild metric:
\begin{equation} \label{e6}
    f(r) = 1 - \frac{2m}{r} \left ( 1 + \frac{4\alpha m^{2}}{L_{*}^{2}}\right ).
\end{equation}
The event horizon can be sought off when $f(r)=0$:
\begin{equation} \label{e7}
    r_{\text{h}} = 2m + \frac{8\alpha m^{3}}{L_{*}^{2}}.
\end{equation}
In Ref. \cite{Mureika_2019}, the effect of EUP correction to the black hole's event horizon, particle and photon orbits, and the temperature was studied. For $\alpha=1$ and $L_*$ ranging from $10^{12}$ m to $10^{14}$ m, the EUP correction becomes relevant for supermassive black holes within the masses of $m=10^9 M_\odot$ to  $m=10^{11} M_\odot$ where $M_\odot$ is the geometrized mass of the Sun if used as an example. Furthermore, $L_*$ may also have direct implication to galactic dynamics or cosmological models \cite{Dabrowski_2019b} if the Hubble length scale is used.

Before proceeding, we comment on some subtlety when plotting equations like Eq. \eqref{e7} due to the EUP term. It is a common practice to transform $r_\text{h}$ into a dimensionless quantity by dividing it by the geometrized mass $m$ to attain brevity in plotting (do not confuse $m$ to the unit of length in meters, m). However, the EUP term in Eq. \eqref{e7} still contains the mass $m$, which cannot be set to unity. Indeed, one can easily verify that using the geometrized value of $m$ leads to the correct horizon radius. For this reason, we will avoid saying $m=1$ on this paper to avoid confusion.

We now envelope the EUP black hole with dark matter proposed recently by Konoplya  \cite{Konoplya_2019}. The mass of the black hole is introduced as a piece-wise function imposing three domains:
\begin{align} \label{e8}
    \mathcal{M}(r)=\begin{cases}
    m_{\text{EUP}}, & r<r_{s};\\
    m_{\text{EUP}}+ M \mathcal{G}(r), & r_{s}\leq r\leq r_{s} + \Delta r_{s};\\
    m_{\text{EUP}}+ M, & r>r_{s}+\Delta r_{s}
    \end{cases}
\end{align}
where
\begin{equation} \label{e9}
    \mathcal{G}(r)=\left (3-2\frac{r-r_{s}}{\Delta r_{s}}\right)\left ( \frac{r-r_{s}}{\Delta r_{s}} \right )^{2}.
\end{equation}
\begin{figure} [htpb]
    \centering
    \includegraphics[width=\columnwidth]{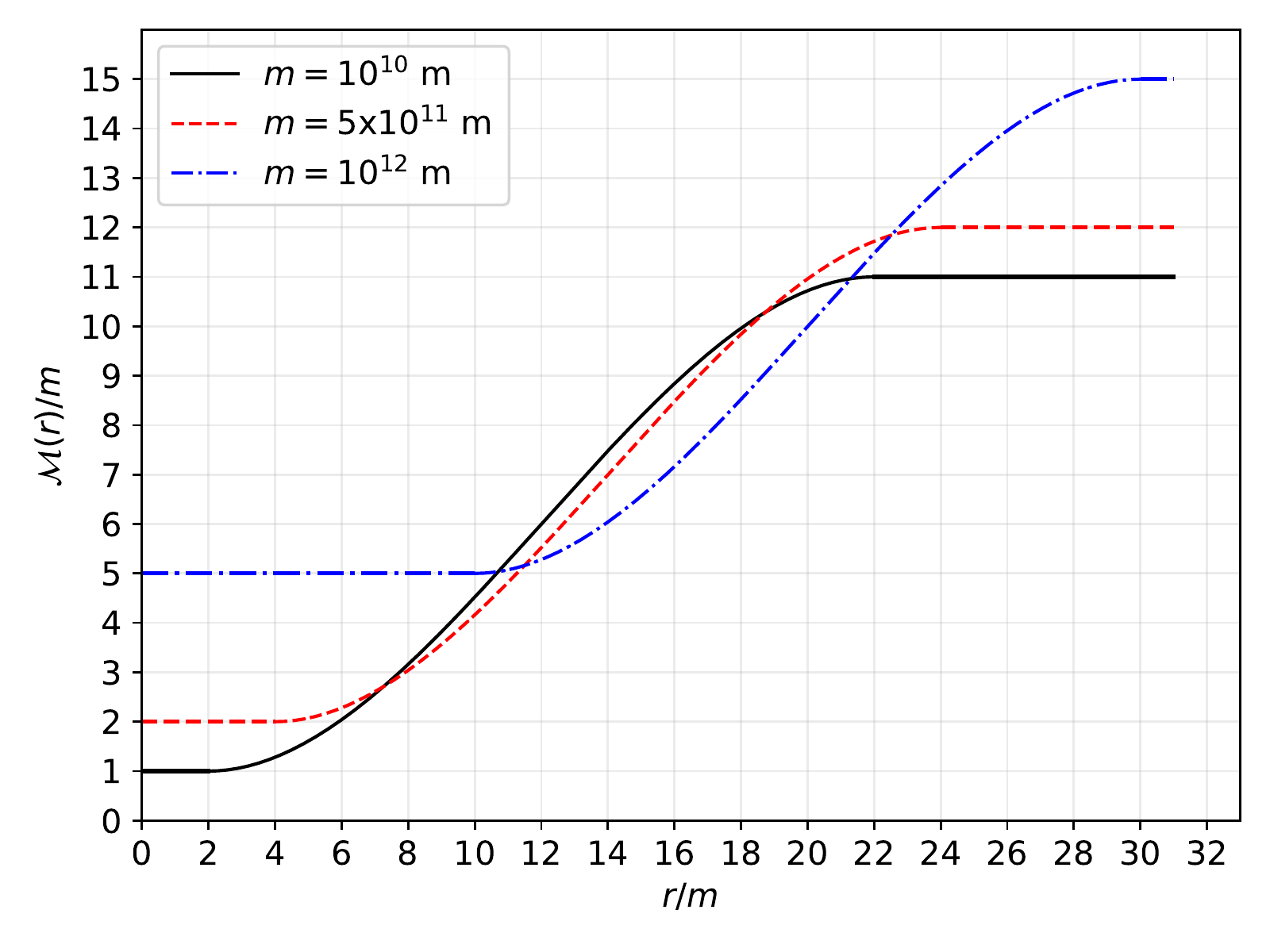}
    \caption{Example of a choice for mass function $\mathcal{M}(r)$ when $L_*=10^{12}$ m. Here, $M=10m$, $r_{\text{h}} = r_s$, and $\Delta r_{s}=20m$. We can see that $\mathcal{M}(r)$ behaves differently for different black hole mass $m$.}
    \label{fig1}
\end{figure}
Here, $m_{\text{EUP}}$ is given by Eq. \eqref{e5}, $ M $ is the dark matter mass, $ r_{s} $ is the dark matter inner shell radius, $ \Delta r_{s} $ is the thickness of the dark matter. It is clear, especially when looking at the third domain, that the dark matter mass $M$ is modeled as an invisible additional effective mass to the black hole, making the mass function less model-dependent \cite{Konoplya_2019}. While we focus our attention to $M>0$, $M<0$ represents the negative energy density of matter. Fig. \ref{fig1} shows the plot of the mass function for several black hole masses. Aside from the plot not showing any discontinuity, it also gives information on the horizon radius of the EUP black hole. For $L_*=10^{12}$ m, very large deviation to the horizon radius is already manifested for black hole masses of $m=5$x$10^{11}$ m and beyond.

\begin{figure} [htpb] 
    \centering
    \includegraphics[width=\columnwidth]{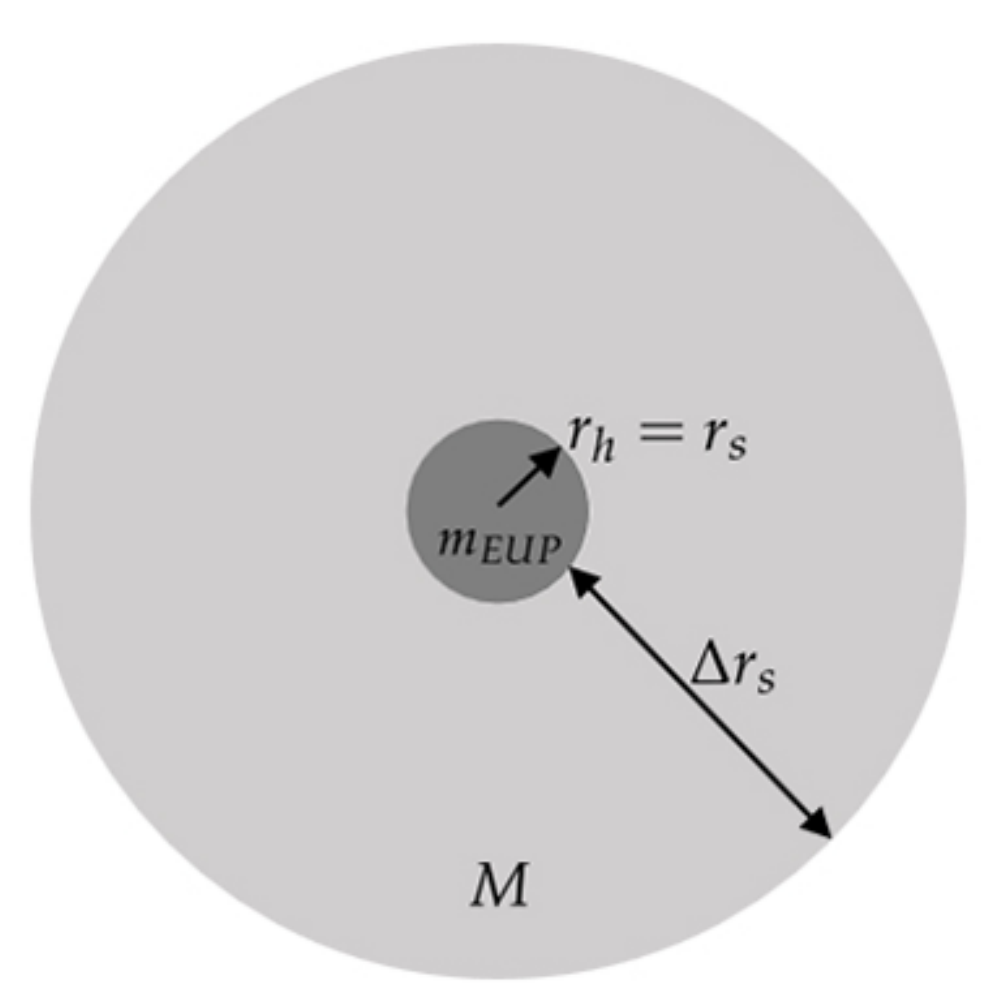}
    \caption{EUP inspired Schwarzschild black hole surrounded by dark matter. Here, the event horizon $r_{\text{h}}$ of the EUP black hole coincides with the inner radius $r_{s}$ of the dark matter shell.}
    \label{fig2}
\end{figure}

Of the three domains, we are especially interested in the non-trivial second domain for it reveals the observed phenomena by some observer inside the dark matter shell. We emphasize too that although $r_s$ may be arbitrary, we restrict our attention to the fact that the inner radius of the dark matter is exactly at $r_{\text{h}}$, which is the event horizon of the EUP black hole given in Eq. \eqref{e7}. See Fig. \ref{fig2}.

Based on Fig. \ref{fig2}, it is easy to see how we may combine the EUP metric function to the dark matter mass. Thus, the static and spherically symmetric spacetime metric
\begin{equation} \label{e10}
    ds^{2} = -f(r) dt^{2} + f(r)^{-1} dr^{2} + r^{2} d\theta ^{2} + r^{2} sin^{2} \theta \; d\phi^{2}
\end{equation}
can have the metric function $ f(r) $ as
\begin{align} \label{e11}
    f(r) &= 1 - \frac{2\mathcal{M}(r)}{r}=1-\frac{2}{r}\biggl[m\left(1 + \frac{4\alpha m^{2}}{L_{*}^{2}}\right) \nonumber \\
    &+M\left (3-2\frac{r-r_{s}}{\Delta r_{s}}\right)\left ( \frac{r-r_{s}}{\Delta r_{s}} \right )^{2}\biggr]
\end{align}

Due to Eq. \eqref{e11}, the EUP-Schwarzschild metric is now affected by dark matter mass as one form of an astrophysical environment. Such type of black hole is called a "dirty" black hole, which first originated from Visser in 1992 \cite{PhysRevD.46.2445}. A dirty black hole may come from a toy model or a metric that is derived from a configuration coming from empirical data. Although generic, they possess sufficient generality \cite{Nielsen2019} that gives notable insights into the effect of a particular astrophysical environment on the geometry of the black hole.

\subsection{Horizon radius}
\begin{figure} [htpb] 
    \centering
    \includegraphics[width=\columnwidth]{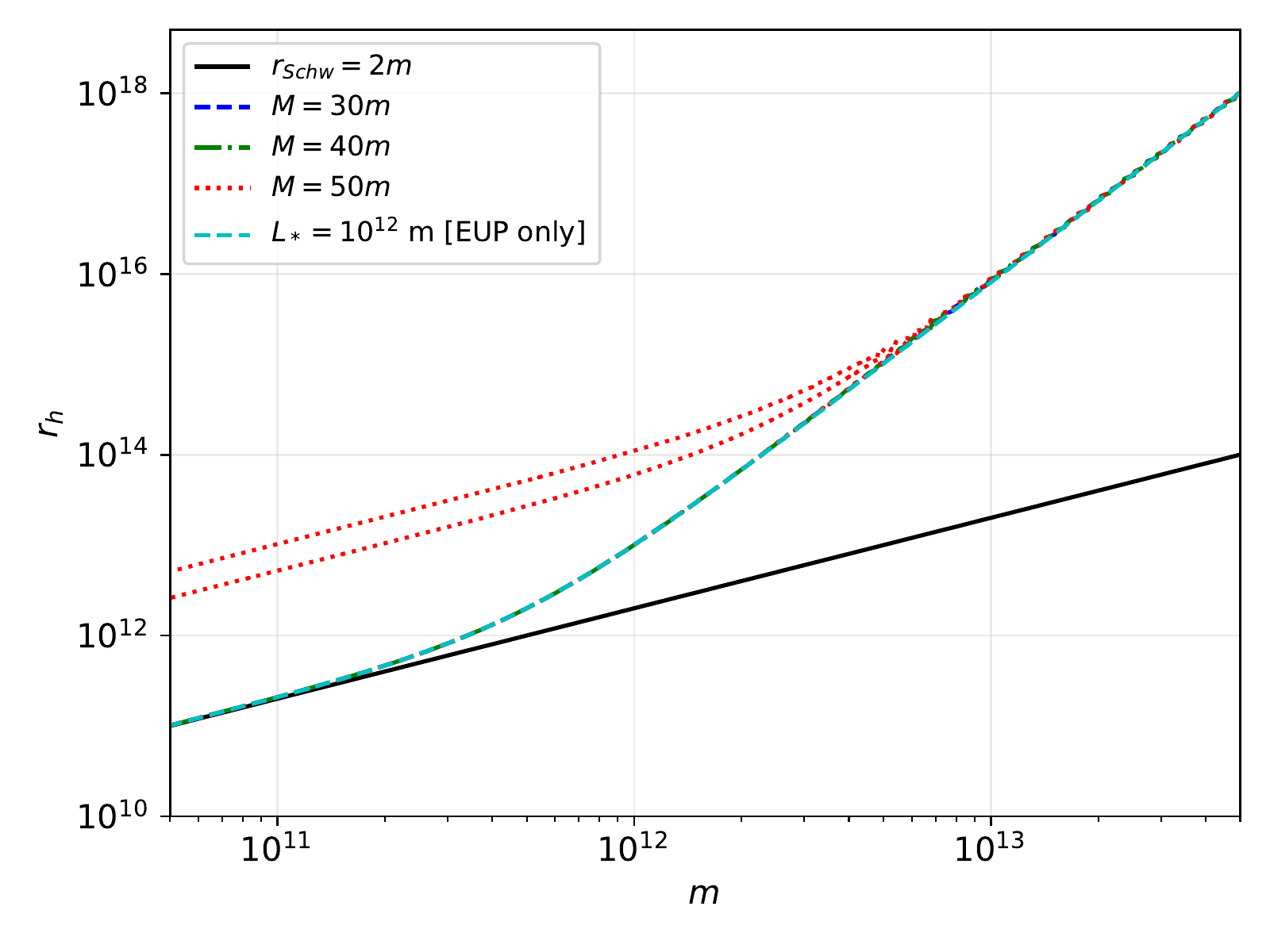}
    \caption{Horizon radius $r_{\text{h}}$ vs. black hole mass $m$. Here, $\Delta r_s=100m$, $L_*=10^{12}$ m. The plot compares the location of $r_{\text{h}}$ between the Schwarzschild BH, EUP BH with dark matter, and the EUP BH alone (last label in the legend).}
    \label{fig3}
\end{figure}

Following standard prescription, the radius of the event horizon $r_{\text{h}}$ can be found by setting $f(r)=0$. For the Schwarzschild black hole with EUP, $r$ can be easily solved analytically. However, when the EUP black hole is surrounded with dark matter as described by Eq. \eqref{e8}, solving for $r$ can be fairly complicated when $r_s \neq r_{\text{h}}$. To simplify the analysis, we set $r_s = r_{\text{h}}$ and solving for $r$ readily gives the horizon of the EUP black hole in Eq. \eqref{e7}. Fig. \ref{fig3} shows the logarithmic plot for location of $r_{\text{h}}$ for a given dark matter shell thickness $\Delta r_s$, and large fundamental length scale $L_*$.
From Fig. \ref{fig3}, the line plot for the Schwarzschild black hole's event horizon is as expected. However, the event horizon of an EUP black hole with $L_* = 10^{12}\text{m}$ begins to deviate from the Schwarzschild one near $m=10^{11}\text{m}$ $\sim$ $4.50$x$10^{30}$ kg, and this deviation starts to get noticeable as the black hole mass further increases. When the EUP black hole is surrounded by the dark matter of mass of $M=30m$ and $M=40m$, the horizon radius is identical to that of the EUP black hole, which is expected since $r_s = r_{\text{h}}$. Three horizons are formed when $M=50m$, which tells us that such a parameter has no physical significance, at least for a black hole mass of $m\sim5$x$10^{12}$ m and below.

\subsection{Effective potential and ISCO of a massive particle}
\begin{figure} [htpb] 
    \centering
    \includegraphics[width=\columnwidth]{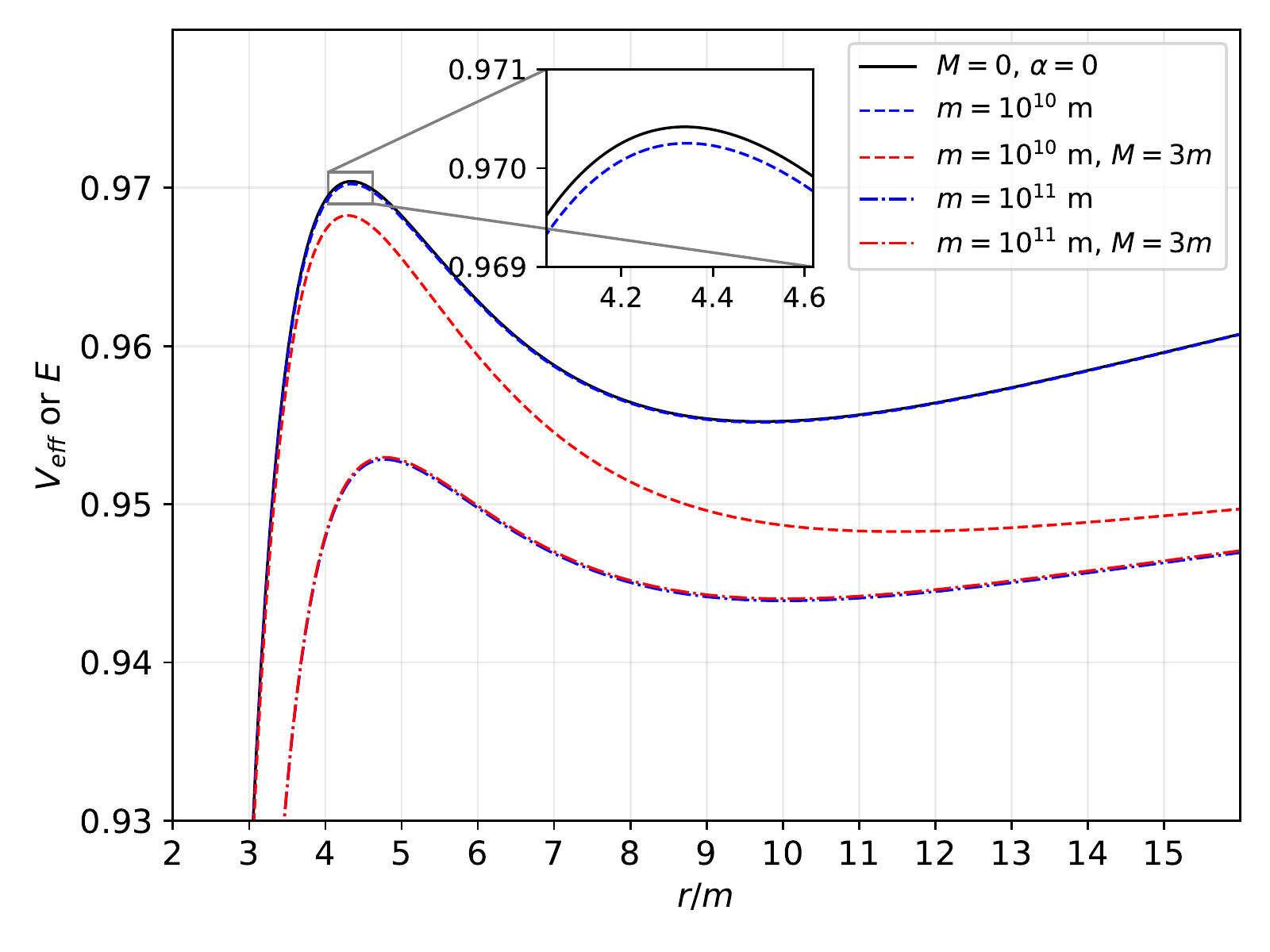}
    \caption{Effective potential $V_{\text{eff}}$ vs. radial distance $r/m$. Here, the chosen value for the large fundamental length scale is $L_*=10^{12}$ m, and $l=3.75$ for the particle's angular momentum.}
    \label{nfig4}
\end{figure}

We now turn our attention to time-like orbits. To do so, we use the following Lagrangian  \cite{chandrasekhar1985mathematical},
\begin{equation} \label{e12}
\mathcal{L}=\frac{1}{2}g_{\mu \nu }\dot{x}^{\mu }\dot{x}^{\nu },
\end{equation}
where $\dot{x}^{\mu }=u^{\mu }=dx^{\mu }/d\lambda $ in which $u^{\mu }$ is the particle's four-velocity with the affine parameter $\lambda $. The affine parameter is defined in terms of the proper time $\tau=\mu\lambda$ where $\mu$ is proportional to the particle's unit rest mass. This affine parameter is chosen so that it can accommodate both time-like and null-orbits. Noting the coordinate independence of the metric in $t$ and $\phi$, we find the conserved quantities
\begin{equation} \label{e13}
E=p_{t}=\frac{\partial \mathcal{L}}{\partial \dot{t}}=g_{tt}\dot{t},\quad l=-p_{\phi }=-\frac{\partial \mathcal{L}}{\partial \dot{
\phi}}=-g_{\phi \phi }\dot{\phi},
\end{equation}
where $E$ and $l$ are the particle's energy and angular momentum per unit mass respectively. In particular, in terms of the initial velocity $v_o$ of the particle, it is well-known that $E=1/\sqrt{1-v_o^2}$. By using the relation $bv_o=l/E$, which is consistent to that of photons impact parameter when $v_o=1$, the angular momentum can be expressed as $l=bv_o/\sqrt{1-v_o^2}$ \cite{He2020}.
Substituting Eq. \eqref{e13} to the normalized four-velocity of the particle
\begin{equation} \label{e14}
-1=g_{\mu \nu }\dot{x}^{\mu }\dot{x}^{\nu},
\end{equation}
we get the radial equation as
\begin{equation} \label{e15}
\dot{r}^2=E^2-f(r)\left(1+\frac{l^{2}}{r^2}\right).
\end{equation}
The second term in Eq. \eqref{e15} is the effective potential of the massive particle:
\begin{equation} \label{e17}
V_{\text{eff}}^2=f(r)\left(1+\frac{l^2}{r^2}\right),
\end{equation}
where one can find the critical angular momentum of a massive particle in orbit if one satisfies the conditions
\begin{equation} \label{e18}
V_{\text{eff}}=\left. \frac{\partial V_{\text{eff}}}{\partial r}\right\vert
_{r=r_{o}}=0,
\end{equation}
which can be easily calculated for the Schwarzschild and EUP black holes. However, when dark matter mass $M$ is introduced, the resulting equation can be fairly cumbersome. For such a case, we can do a numerical plot of $V_{\text{eff}}$ to analyze qualitatively the particle orbits and this is shown in Fig. \ref{nfig4}.
Note that the positive root is chosen in Eq. \eqref{e17} since negative energies for a massive particle are not allowed in a non-rotating black hole. Furthermore, since we are just describing the qualitative picture of the different orbits, we can specify an arbitrary value for the angular momentum $l$. Thus, for a given value of impact parameter $b$, one can determine the particle's initial velocity. In Fig. \ref{nfig4}, we see how the potential curve is affected when the mass $m$ increases for a given $L_*$. Relative to the Schwarzschild case (solid black line), the EUP correction decreases slightly the peak energy of a particle in the unstable circular orbit (see inset plot). If the EUP black hole is surrounded by the dark matter mass of $M=3m$, the peak gets lower as shown in the red dashed curve. Further increasing $m$ decreases the peak energy, and the same effect can be said if it has dark matter. We can tell that the effect of dark matter diminishes when the mass of the EUP black hole increases. The plot also shows how the energy required for a particle to have stable circular orbit decreases, as well as how its radius varies from the center of the EUP black hole. When it comes to bound elliptical orbits, comparing the blue and red dashed lines indicates that dark matter increases drastically the aphelion radius.

\begin{figure} [htpb] 
    \centering
    \includegraphics[width=\columnwidth]{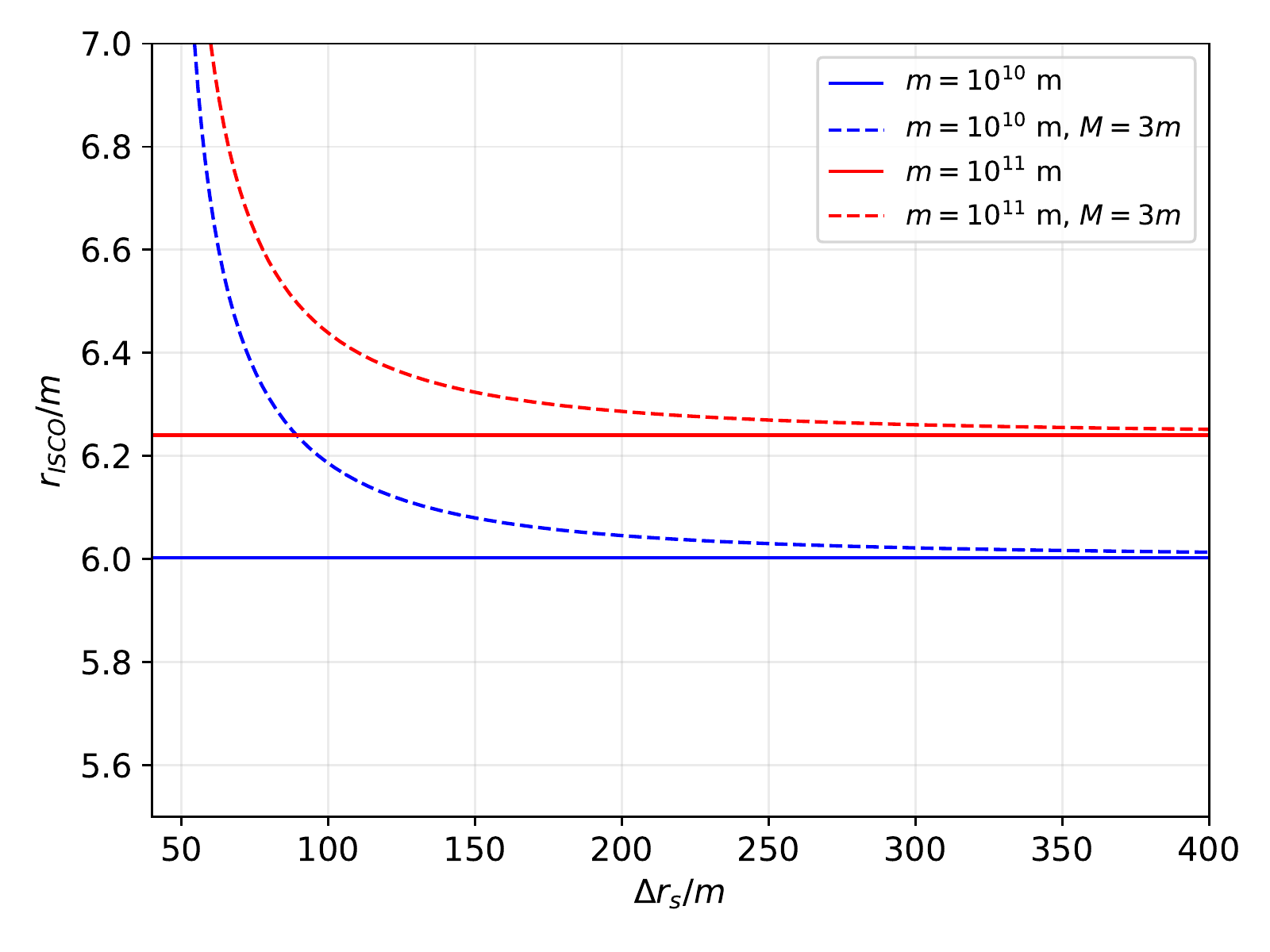}
    \caption{Location of ISCO radius $r_{\text{ISCO}}$ vs. dark matter thickness $\Delta r_s$. Here, the large scale fundamental length is $L_*=10^{12}$ m. The plot compares the $r_{\text{ISCO}}$ between EUP BH and EUP with dark matter.}
    \label{fig4}
\end{figure}

The location of the ISCO radius is also worth investigating due to its importance to the dynamics of the accretion disk. To locate the ISCO radius, we can use the radial equation in Eq. \eqref{e15} and rewrite is as
\begin{equation} \label{e19}
\mathcal{R}(r)=r^4\dot{r}^2=E^{2}r^4-r^2f(r)\left(r^2+l^2\right)
\end{equation}
and the energy for circular orbit can be obtained by solving simultaneously the resulting equations from the condition
\begin{equation} \label{e20}
\mathcal{R}(r)=\left. \frac{\partial \mathcal{R}(r)}{\partial r}\right\vert
_{r=r_{o}}=0.
\end{equation}
The result is
\begin{equation} \label{e21}
    E^2_{\text{circ}}=\frac{2f(r)}{(2f(r)-f'(r)r)}.
\end{equation}
For the energy required in ISCO radius, we require another condition that
\begin{equation} \label{e21n}
    \left. \frac{\partial^2 \mathcal{R}(r)}{\partial r^2}\right\vert_{r=r_{o}}=0
\end{equation}
as it would mean that the maxima and minima (for example in Fig. \ref{nfig4}) will coincide. Using Eq.\eqref{e21n}, we find that:
\begin{equation} \label{e22}
    E^2_{\text{ISCO}}=\frac{r^2f(r)f''(r)-2r^2f'(r)^2-5rf(r)f'(r)-8f(r)^2}{2r^2f''(r)+2rf'(r)-8f(r)}.
\end{equation}
Since Eqs. \eqref{e21} and \eqref{e22} are equal, we can solve $r$ numerically via
\begin{equation} \label{e23}
    rf(r)f''(r)-2rf'(r)^2+3f(r)f'(r)=0.
\end{equation}
Fig. \ref{fig4} shows the location of ISCO radius. It is well-known that for Schwarzschild black hole, the ISCO radius is $6m$. As shown in the plot, the EUP correction in $m=10^{10}$ m slightly deviates from $6m$. Noticeable deviation occurs for $m=10^{11}$ m. It is also clear how the slight addition of dark matter to the EUP black hole further increases the radius. Drastic increase to the radius is also gleaned when $\Delta r_s$ gets smaller. As the dark matter density decreases, the dashed lines approximates to the EUP ISCO radius. To close this subsection, we comment that the behavior of massive particles, as seen in Figs. \ref{nfig4} and \ref{fig4}, is easily affected by dark matter even at low mass $M$.

\subsection{Temperature}
We can use the well-known formula for the Hawking radiation \cite{Ovgun_2018} to calculate the temperature of EUP black hole with dark matter:
\begin{align} \label{e24}
    T&=\frac{1}{4\pi}f'(r) = \frac{1}{4\pi L_*^{2}r^{2}} \biggl[8\alpha m^{3}+L_*^{2}\biggl\{ 2m \nonumber \\
    &+M\biggl[\frac{6r_s^{2}-6r^{2}}{\Delta r_s^{2}}+\frac{8}{\Delta r_s^{3}}\left(r+\frac{r_s}{2}\right)(r-r_s)^{2}\biggr]\biggr\}\biggr].
\end{align}
Since $r$ is evaluated at $r_{\text{h}}$, and $r_s=r_{\text{h}}$, an observer anywhere inside the dark matter shell will measure the Hawking temperature to be the same as that of the EUP black hole alone:
\begin{equation} \label{e25}
    T=\frac{L_{*}^{2}}{8\pi m\left(4\alpha m^{2}+L_{*}^{2}\right)}.
\end{equation}
We see that the Hawking temperature is independent of the dark matter mass $M$, as far as the second domain in Eq. \eqref{e8} is concerned. For the case that if $m<<L_*$,
\begin{equation} \label{e26}
    T\sim\frac{1}{8\pi m}-\frac{\alpha m}{2\pi L_{*}^{2}}
\end{equation}
and if $m>>L_*$,
\begin{equation} \label{e27}
    T\sim\frac{L_{*}^{2}}{32\pi \alpha m^3}.
\end{equation}

\section{Shadow of EUP black hole surrounded by dark matter} \label{sec3}
We compute for the radius of the photon sphere following the method presented in Perlick et al.  \cite{Perlick_2015}. Without loss of generality, we analyze the null geodesic in the equatorial plane only such that $\theta = \pi/2$. For a spherically static and symmetric (SSS) spacetime, the Hamiltonian for light rays is in general given by
\begin{equation} \label{e28}
    H = \frac{1}{2} g^{ik} p_{i} p_{k} = \frac{1}{2} \left( -\frac{p_{t}^{2}}{A(r)} + \frac{p_{r}^{2}}{B(r)} + \frac{p_{\phi }^{2}}{D(r)} \right),
\end{equation}
where $A(r)=f(r)$, $B(r)=f(r)^{-1}$, and $D(r)=r^2$ wherein the metric function $f(r)$ is given in Eq. \eqref{e11}. The equations of motion for null particles are then
\begin{equation} \label{e29}
    \dot{x}^{i} = \frac{\partial H}{\partial p_{i}}, \quad \quad \dot{p}_{i} = -\frac{\partial H}{\partial x^{i}}.
\end{equation}
Here, $\dot{x}=dx/d\lambda$ and $\dot{p}$ represents the conjugate momenta. Eq. \eqref{e29} gives
\begin{equation} \label{e30}
    \dot{t} = -\frac{p_{t}}{f(r)}, \quad \quad \dot{\phi } = \frac{p_{\phi }}{r^2}, \quad \quad \dot{r} = -\frac{p_{r}}{f(r)^{-1}},
\end{equation}
$$\dot{p}_{t} = 0, \quad \quad \dot{p}_{\phi } = 0,$$
\begin{equation} \label{e31}
     \dot{p}_{r} = \frac{1}{2} \left( -\frac{p_{t}^{2} f'(r)}{f(r)^{2}} + \frac{p_{r}^{2} f'(r)^{-1}}{f(r)^{-2}} + \frac{2p_{\phi }^{2}}{r^{3}} \right).
\end{equation}
Setting $ H = 0 $, we have
\begin{equation} \label{e32}
    -\frac{p_{t}^{2}}{f(r)} + \frac{p_{r}^{2}}{f(r)^{-1}} + \frac{p_{\phi }^{2}}{r^2} = 0,
\end{equation}
and it now follows that
\begin{equation} \label{e33}
    \frac{dr}{d\phi } = \frac{\dot{r}}{\dot{\phi }} = \frac{r^2} {f(r)^{-1}}\frac{p_r}{p_{\phi}}.
\end{equation}
Setting $p_t=-\omega_{o}$, and using $p_r$, we can get the relation how $r$ changes with $\phi$:
\begin{equation} \label{e34}
    \frac{dr}{d\phi } = \pm \frac{r}{f(r)^{-1/2}} \sqrt{\frac{\omega _{o}^{2}}{p_{\phi }^{2}} h(r)^{2} - 1},
\end{equation}
where 
\begin{equation} \label{e35}
    h(r)^{2} = \frac{r^2}{f(r)}
\end{equation}
is defined. For a circular light orbit, the radial velocity and acceleration should be $ \dot{r} = 0 $ and $ \ddot{r} = 0 $ respectively, and hence, $ p_{r} = 0 $. Eq. \eqref{e32} then becomes
\begin{equation} \label{e36}
    0 = -\frac{\omega _{o}^{2}}{f(r)} + \frac{p_{\phi }^{2}}{r^2}.
\end{equation}
Since $ \dot{p}_{r} = 0 $, Eq. \eqref{e31} can be rewritten as
\begin{equation} \label{e37}
    \dot{p}_{r} = 0 = -\frac{\omega _{o}^{2} f'(r)}{f(r)^{2}} + \frac{2p_{\phi }^{2}}{r^3}.
\end{equation}
Using Eqs. \eqref{e36} and \eqref{e37}, we find
\begin{equation} \label{e38}
    p_{\phi }^{2} = r^2  \frac{\omega _{o}^{2}}{f(r)},
\end{equation}
\begin{equation} \label{e39}
    p_{\phi }^{2} = r^3  \frac{\omega _{o}^{2} f'(r)}{f(r)^{2}}.
\end{equation}
The implication of subtracting Eqs. \eqref{e38} and \eqref{e39} gives the information how to find the radius of the photon sphere:
\begin{equation} \label{e40}
    0 = \frac{d}{dr} h(r)^{2},
\end{equation}
which leads us to the expression
\begin{equation} \label{e41}
    2\alpha m^{3}+L_*^{2}\biggl\{\frac{m}{2}-\frac{r}{6}+\frac{M(r-r_s)}{\Delta r_s^{2}}\biggl[\frac{r_s(r-r_s)}{\Delta r_s}+\frac{r}{2}-\frac{3r_s}{2}\biggr]\biggr\} =0.
\end{equation}
The above can be solved for $r$ to determine the photon sphere radius $r_{\text{ph}}$. The result is
\begin{equation} \label{e42}
    r_{\text{ph}}=\frac{\pm\sqrt{K}+L_*\left(\frac{12Mr_s^{2}}{\Delta r_s^{3}}+\frac{12Mr_s}{\Delta r_s^{2}}+1\right)}{L_*\left(\frac{12Mr_s}{\Delta r_s^{3}}+\frac{6M}{\Delta r_s^{2}}\right)},
\end{equation}
where
\begin{align} \label{e43}
    K&=\frac{36L_*^{2}M^{2}r_s^{2}}{\Delta r_s^{4}}-\frac{72M}{\Delta r_s^{2}}\biggl\{ \frac{r_s}{\Delta r_s}\biggl[4\alpha m^{3}+L_*^{2}\left(m-\frac{r_s}{3}\right)\biggr] \nonumber \\
    &+2\alpha m^{3}+L_*^{2}\left(\frac{m}{2}-\frac{r_s}{3}\right)\biggr\} +L_*^{2}
\end{align}
Here, we choose the lower sign since the upper sign represents a photon sphere that is far from the black hole. We see that Eq. \eqref{e42} is a complicated equation for $r_\text{ph}$ due to the second domain of the mass function in Eq. \eqref{e9}. However, we can do a numerical analysis to locate the corresponding photon sphere radius as we compare the EUP black hole and the one with dark matter.
\begin{figure} [htpb] 
    \centering
    \includegraphics[width=\columnwidth]{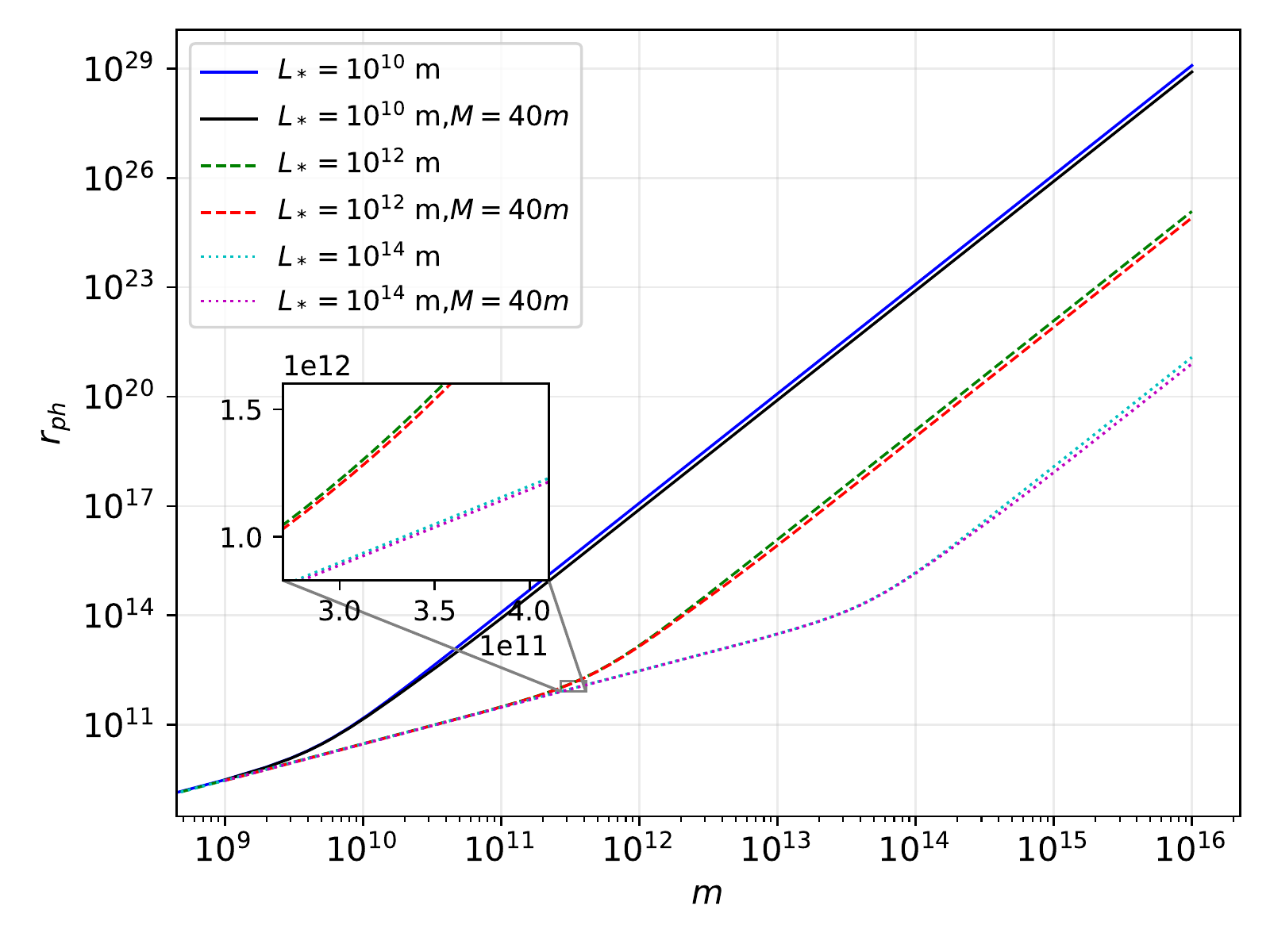}
    \caption{Photon sphere radius $r_{\text{ph}}$ vs. black hole mass $m$. Here, $\Delta r_{s}=100m$. The plot shows how $r_{\text{ph}}$ changes due to the influence of the EUP alone, and EUP with dark matter.}
    \label{fig5}
\end{figure}
Looking at Fig. \ref{fig5}, we see that the effect of EUP alone is to increase the radius of the photon sphere for a given black hole mass $m$ where this increase depends on the large scale parameter $L_*$. For example, when $L_*=10^{10}$ m, large deviation to $r_{\text{ph}}$ immediately begins to less massive black holes. As for the dark matter effect on the EUP BH, generally, it decreases the value of $r_{\text{ph}}$. Minuscule effect due to dark matter can be seen as $r_{\text{ph}}$ begins to deviate due to the EUP correction (see inset plot). However, when this deviation gets large (for more massive BHs), the dark matter effect becomes more noticeable. Furthermore, a large amount of dark matter mass is needed to see such deviation.

As for the black hole shadow, it depends on the initial direction of light rays that spiral towards the outermost photon sphere. The angular radius of the shadow $ \alpha _{\text{sh}} $ is defined by
\begin{equation} \label{e44}
    \cot(\alpha_{\text{sh}}) = \frac{f(r)^{-1/2}}{r} \left. \frac{dr}{d\phi } \right|_{r=r_{o}}.
\end{equation}
If the light ray goes out again after reaching $ r_{\text{ph}} $, the orbit equation in Eq. \eqref{e34} can be rewritten as
\begin{equation} \label{e45}
    \frac{dr}{d\phi } = \pm \frac{r}{f(r)^{-1/2}} \sqrt{\frac{h(r)^{2}}{h(r_{\text{ph}})^{2}} - 1}.
\end{equation}
Thus, the angular radius of the shadow becomes
\begin{equation} \label{e46}
    \cot^{2}(\alpha _{sh}) = \frac{h(r_{o})^{2}}{h(r_{\text{ph}})^{2}} - 1,
\end{equation}
and by using a trigonometric identity, $ 1 + \cot^{2}(\alpha_{\text{sh}}) = \frac{1}{\sin^{2}(\alpha_{\text{sh}})} $, it can be rewritten as
\begin{equation} \label{e47}
    \sin^{2}(\alpha _{sh}) = \frac{h(r_{\text{ph}})^{2}}{h(r_{o})^{2}},
\end{equation}
where $ h(r_{\text{ph}})$ must be evaluated using the photon sphere radius in Eq. \eqref{e42}, and $ r_{o} $ specifies the location of the observer which is usually $r_o\rightarrow\infty$. The radius of the shadow of EUP black hole surrounded by dark matter can then be calculated as
\begin{align} \label{e48}
    R_{\text{sh}}&=r_o\sin\alpha_{\text{sh}}=r_{\text{ph}}\biggl(1-\frac{2}{r_{\text{ph}}}\biggl\{ m\biggl(\frac{4\alpha m^{2}}{L_*^{2}}+1\biggr) \nonumber \\
    &+\frac{M(r_{\text{ph}}-r_s)^{2}}{\Delta r_s^{2}}\biggl[3-\frac{2(r_{\text{ph}}-r_s)}{\Delta r_s}\biggr]\biggr\} \biggr)^{-1/2}
\end{align}
Again, we see that Eq. \eqref{e48} is a cumbersome expression if we substitute $r_{\text{ph}}$ in Eq. \eqref{e42}. However, for a given dark matter mass $M$, we can do an approximation in which $\Delta r_s$ is very large in comparison to $M$. It means that dark matter is highly diluted. The result is
\begin{equation} \label{e49}
    R_{\text{sh}}=3\sqrt{3}m+\frac{9\sqrt{3}m^{2}M}{\Delta r_s^{2}}+\frac{12\sqrt{3}\alpha m^{3}}{L_*^{2}}+\frac{72\sqrt{3}\alpha m^{4}M}{L_*^{2}\Delta r_s^{2}},
\end{equation}
where we note that the 4th term can be vanishingly small. It is enough to consider the first three terms where both dark matter mass and EUP correction are present and contributes to any change in the known Schwarzschild shadow radius of $3\sqrt{3}m$.
\begin{figure} [htpb] 
    \centering
    \includegraphics[width=\columnwidth]{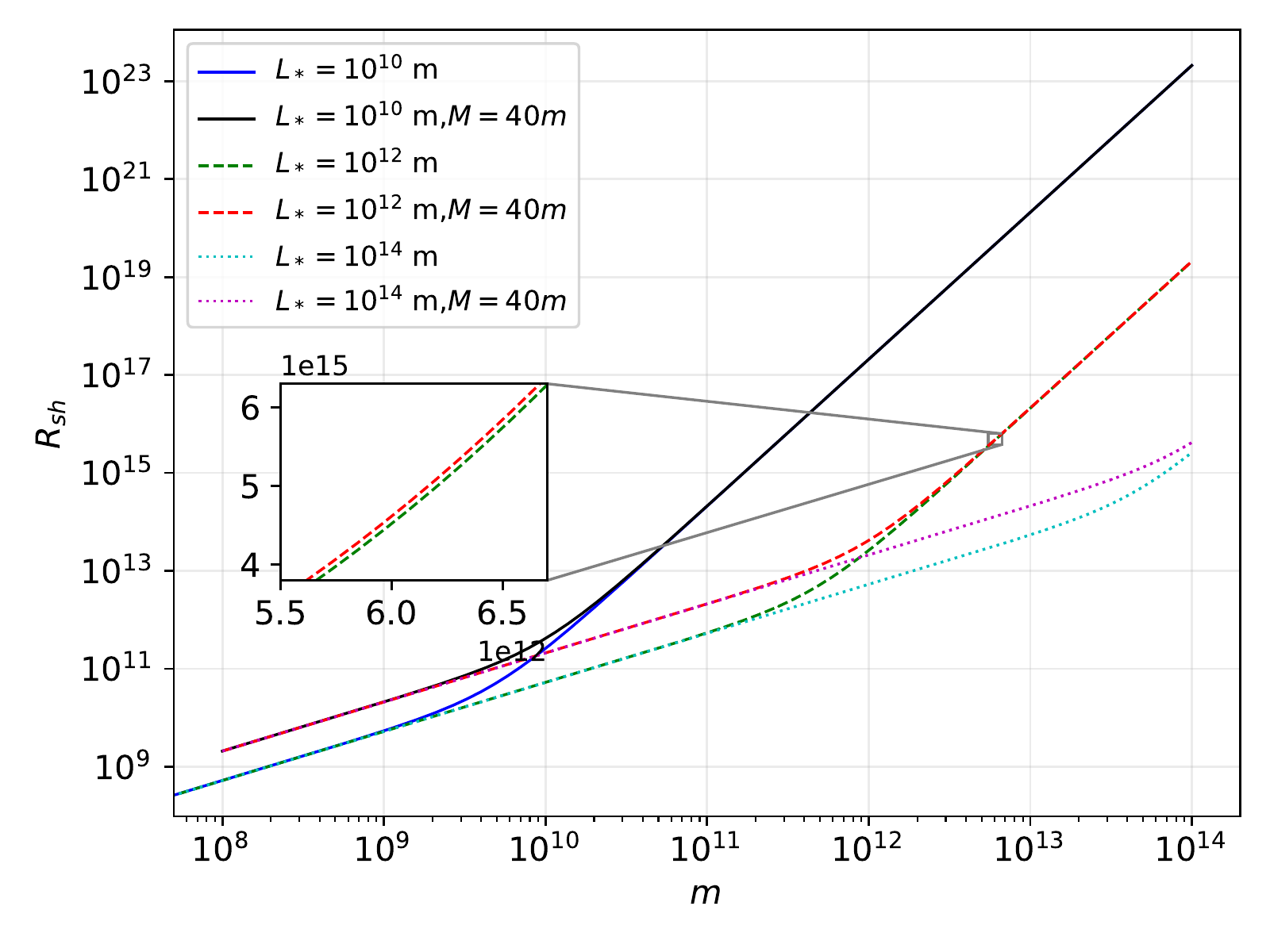}
    \caption{Shadow radius $R_{\text{sh}}$ vs. black hole mass $m$. Here, $\Delta r_{s}=100m$. The plot shows how $R_{\text{sh}}$ changes due to the influence of the EUP alone, and EUP with dark matter. For Sgr. A*, $m=4.3$x$10^{6}$}
    \label{fig6}
\end{figure}
In Fig. 7, let us examine first the effect of $L_*$ alone. One can fairly see that the three given values for $L_*$ have a negligible effect on the shadow radius for low mass black holes ($<10^9$). Deviation begins to manifest first for $L_*=10^{10}$ m, but such a deviation becomes abnormally large for larger black hole mass. Thus, for this value of $L_*$, the deviations which make sense may fall only around the mass range of $10^9-10^{10}$ m. For $L_*=10^{12}$ m, the mass range is around $10^{11}-10^{12}$ m, and finally for $L_*=10^{14}$ m, it is $10^{13}-10^{15}$ m. As for the dark matter effect, it tends to increase the shadow radius relative to the parameter $L_*$. This is in contrast with Fig. \ref{fig5} where the radius of the photon sphere decreases. Based on these results, we have seen two origins for the deviation in $R_{sh}$: one is from the mass of the black hole itself, and another is from the dark matter astrophysical environment. As a final note, we also see in Fig. \ref{fig6} how the dark matter effects delay the EUP effect. For example, when $L_*=10^{12}$ m and without dark matter effects, the deviation occurs midway between $10^{11}-10^{12}$ m. However, due to the dark matter environment, The EUP effect now occurs near black hole masses of around $10^{12}$ m.

\begin{table*}
\scriptsize
\centering
\begin{tabular}{|l|p{1.6cm}|p{1.67cm}|p{1.67cm}|p{1.67cm}|p{1.6cm}|p{1.6cm}|p{1.6cm}|}
\hline
\multicolumn{8}{|c|}{Shadow radius [m]}                                                                                                               \\ 
\hline
\multicolumn{1}{|c|}{\multirow{2}{*}{}} & \multirow{2}{*}{Schw. term} & \multicolumn{3}{c|}{DM contribution} & \multicolumn{3}{c|}{EUP contribution}  \\  
\cline{3-8} \\[-1em]
\multicolumn{1}{|c|}{}                  &                             & $L_*=10^{10}$m & $L_*=10^{12}$m & $L_*=10^{14}$m & $L_*=10^{10}$m & $L_*=10^{12}$m & $L_*=10^{14}$m   \\ 
\hline 
Sgr. A*                                  & $3.30004\text{e}(10)$                           & $3.61755\text{e}(-8)$          & $3.61755\text{e}(-8)$          & $3.61755\text{e}(-8)$          & $5.32421\text{e}(10)$          & $5.32421\text{e}(6)$          & $532.42047$            \\
M87                                     & $2.30236\text{e}(13)$                           & $1.24973\text{e}(-13)$          & $1.24973\text{e}(-13)$          & $1.24973\text{e}(-13)$          & $1.80806\text{e}(19)$          & $1.80806\text{e}(15)$          & $1.80806\text{e}(11)$            \\
UGC 7232                                & $7.67452\text{e}(8)$                           & $7.99625\text{e}(-10)$           & $7.99625\text{e}(-10)$          & $7.99625\text{e}(-10)$         & $6.69652\text{e}(5)$          & $66.96523$          & $6.69652\text{e}(-3)$            \\
\hline
\end{tabular}
\caption{Numerical values of the shadow radius obtained from Eq. \eqref{e49}. The dark matter contribution also depends on $L_*$ since $r_s$ is equal to the event horizon of the EUP black hole. Here, we have used the following values: For Sgr. A* \cite{Dearson_2020}, $m=4.3\text{x}10^{6}M_{\odot}$, $M=3.12\text{x}10^{12}M_{\odot}$, $\Delta r_s=290$ kpc; for M87 \cite{Jusufi_2019}, $m=3\text{x}10^{9}M_{\odot}$, $M=2.19\text{x}10^{13}M_{\odot}$, $\Delta r_s=91.2$ kpc; and for UGC 7232 \cite{Jusufi_2019}, $m=1\text{x}10^{5}M_{\odot}$, $M=1.86\text{x}10^{8}M_{\odot}$, $\Delta r_s=0.35$ kpc.}
\label{tab:table1}
\end{table*}

In Table \ref{tab:table1}, the shadow radius of typical black holes is listed along with the dark matter and EUP contributions. We emphasize first that the values in the Schwarzschild term are estimates since the black hole at the center of each galaxy is rotating. We notice in the table that the deviation due to dark matter is indeed negligible even if different values of $L_*$ are used. This is also true even if we have used the data for the core density and radius of the dark matter distribution \cite{Jusufi_2019}. As for the EUP effect, we can see its dependence on the mass of the black hole. For example, if we consider M87 and the possibility of detecting EUP effects experimentally, the choice $L_*=10^{10}$ m can be ruled out by observation due to its unrealistic contribution to the shadow radius. Speculatively speaking, if one discovers that $L_*$ falls around $10^{12}-10^{13}$ m, then the EUP effect for low mass black holes is now irrelevant or at least would be very difficult to detect.

For the shadow radius to be considerably changed by both dark matter mass and EUP correction, we can use Eq. \eqref{e49} to estimate the dark matter thickness that must be fulfilled:
\begin{equation} \label{e50}
    \Delta r_s=\frac{\sqrt{3mM}}{\sqrt{1-\frac{4\alpha m^{2}}{L_*^{2}}}}.
\end{equation}
Note that if there is no EUP correction, it agrees with the result in Ref.  \cite{Konoplya_2019} where $\Delta r_s = 9.18$x$10^{12}$ m if the data for Sgr. A* is used. It was concluded that because $\Delta r_s$ is way too small compared to the actual size of the dark matter ($~290$ kpc), the requirement $\Delta r_s=\sqrt{3mM}$ is not fulfilled to our galaxy. Based on this argument, Eq. \eqref{e50} may still be applicable if one finds a galaxy where the dark matter distribution is concentrated near the supermassive black hole. We also note that under this requirement, $L_*$ should not be any lower than $2m$. With Sgr. A* and M87, these values are $\sim 1.3\text{x}10^{10}$ m and $\sim 1\text{x}10^{13}$ m respectively. Notice that these values are also consistent with the constraints provided in Ref. \cite{Lu_2019}.

In Ref.  \cite{Lu_2019}, their study about the measurements on the separation between the primary and secondary images in the weak deflection lensing and the apparent size of the photon sphere in the strong deflection lensing left an opportunity to impose constrain to $L_*$. In our galaxy, $L_*\sim10^{10}$ m, while for M87 galaxy, $L_*\sim10^{13}$ m. This is consistent with the observations in Table \ref{tab:table1} for the shadow radius. Using these values, we can calculate the improvement portrayed in Eq. \eqref{e50}. In our galaxy, $\Delta r_s = 4.19$x$10^{13}$ m, which is $\sim3.6$ times the previous value due to the introduction of $L_*$. Here, we used $M = 3.12$x$10^{12}M_\odot$, $m = 4.3$x$10^{6}M_\odot$, and $L_* \sim 1.3$x$10^{10}$ m. We see that even with such improvement, Eq. \eqref{e50} is still not met within our galaxy. For M87 galaxy, whose dark matter mass is estimated $95\%$ of its total mass  \cite{Binney_1981,Stewart_1984,Fabricant_1983} $M_{\text{M87}}=3$x$10^{13}M_\odot$ and the central black hole mass to be $m=3$x$10^{9}M_\odot$, we find that $\Delta r_s=7.5$x$10^{14}$ m without EUP correction. With EUP correction, $\Delta r_s=1.6$x$10^{15}$ m, which is an improvement by $\sim1.13$ times the previous value. Unfortunately, $\Delta r_s$ with a EUP correction is still in many orders of magnitude smaller relative to the radius of the galaxy (which is $\sim 5$x$10^{20}$ m for the Milky Way, and $\sim 1.14$x$10^{21}$ m for M87). This is true even if we have used the core density and radius of the dark matter in the M87 galaxy. Thus, we cannot expect the combination of dark matter and EUP correction to manifest itself to the shadow of the central supermassive black hole unless the dark matter distribution is concentrated near the black hole..

\section{Weak Deflection Angle of Light by EUP Black Hole in Dark Matter Halo} \label{sec4}
\begin{figure} [htpb!] 
    \centering
    \includegraphics[width=\columnwidth]{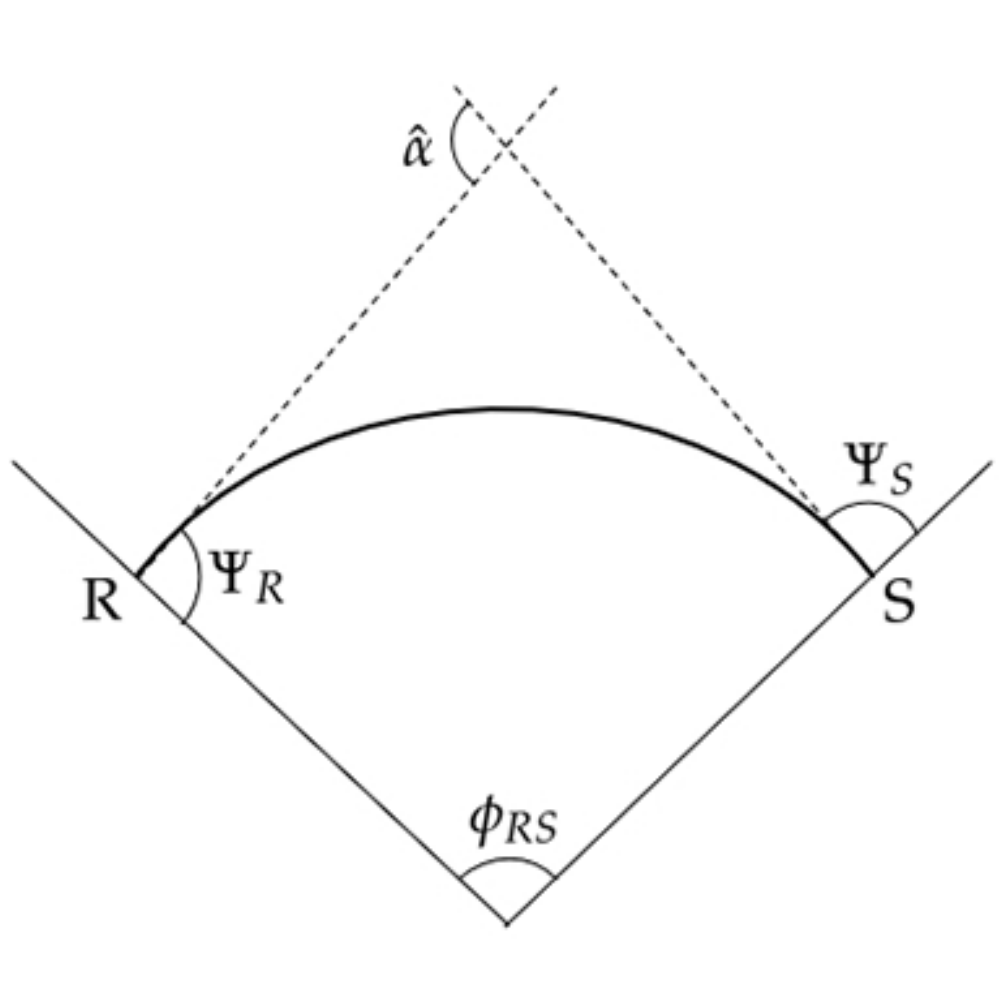}
    \caption{Weak Deflection Angle for Thin Lens Approximation.}
    \label{fig8}
\end{figure}

In this section we calculate the weak deflection angle using the Ishihara et al. method \cite{Ishihara_2016} to see if there will be some improvement to Eq. \eqref{e50} for the dark matter thickness estimate. From the Gauss-Bonnet theorem, there is a generalized correspondence between the weak deflection angle $\hat{\alpha}$ and the surface integral of the Gaussian curvature:
\begin{align} \label{e51}
    \hat{\alpha}&=\phi_{RS}+\Psi_{R}-\Psi_{S} \nonumber \\
    &=\int_{u_{R}}^{u_{o}}\frac{1}{\sqrt{F(u)}}du+\int_{u_{S}}^{u_{o}}\frac{1}{\sqrt{F(u)}}du+\Psi_{R}-\Psi_{S}.
\end{align}
Here, $\phi_{RS}$ is the equatorial angle separation between the source $S$ and receiver $R$, $\Psi$ is the angle measured at the location of the source and receiver. Further, $F(u)$ is the orbit equation as a function of the inverse of the radial coordinate $r$ ($u=1/r$), and $u_o$ is the distance of the closest approach. Eq. \eqref{e51} helps us to find the weak deflection angle of non-asymptotic spacetimes such as those that involve the cosmological constant $\Lambda$. Looking at the mass function in Eq. \eqref{e8}, the model introduces non-asymptotic flatness, and thus, we cannot simply use the asymptotic form of the Gauss-Bonnet theorem. See Fig. \ref{fig8}.

In terms of metric coefficients $A(u)$, $B(u)$, and $D(u)$ (see Sect. \ref{sec3}), the orbit equation is defined as
\begin{align} \label{e53}
    \left(\frac{du}{d\phi}\right)^2 \equiv F(u)&=\frac{u^4D(u)(D(u)-A(u)b^2)}{A(u)B(u)b^2} \nonumber \\
    &=\frac{1/b^2-u^2f(u)}{f(u)f(u)^{-1}}.
\end{align}
Here, we used the substitution $u=1/r$ is used, and $b=l/E$ is the impact parameter. The closest approach $u_o$ can be found by iteratively solving Eq. \eqref{e53} while imposing the boundary condition that $\frac{du}{d\phi}\big|_{\phi=\frac{\pi}{2}}=0$. In doing so, we find the following:
\begin{align} \label{e54}
    F(u) &= \frac{1}{b^{2}} - u^{2} + 2mu^{3}+ \frac{8 \alpha m^{3} u^{3}}{L_{*}^{2}} \nonumber \\
    &+ \frac{6Mu (1-r_{s}u)^{2}}{\Delta r_{s}^{2}} + O \left (\frac{1}{\Delta r_{s}^{3}} \right ),
\end{align}
\begin{equation} \label{e55}
    u_o=\frac{\sin\phi}{b}+\frac{m\left(1+\cos^{2}\phi\right)}{b^{2}}+\frac{4\alpha m^{3}}{b^{2}L_*^{2}}+\frac{3Mr_{s}^{2}}{b^{2}\Delta r_{s}^2}
\end{equation}
which can help us solve the integrals in Eq. \eqref{e51}. Note that to obtain Eq. \eqref{e54}, the approximation $\Delta r_s>>M$ must be imposed. Since $\Delta r_s$ is so large, it will suffice to omit the higher order terms involving $1/\Delta r_s^3$ to simplify the calculation. We also note that in this weak field limit, the leading order in $\alpha$ is coupled with $m^3$. The evaluation of the integral yields $ \phi _{RS} $:
\begin{widetext}
\begin{align} \label{e57}
    \phi_{RS} & = \pi -\arcsin(bu_R)-\arcsin(bu_S)-\frac{m}{b}\left[\frac{\left(b^{2}u_R^{2}-2\right)}{\sqrt{1-b^{2}u_R^{2}}}+\frac{\left(b^{2}u_S^{2}-2\right)}{\sqrt{1-b^{2}u_S^{2}}}\right] 
    -\frac{4\alpha m^{3}}{bL_*^{2}}\left[\frac{\left(b^{2}u_R^{2}-2\right)}{\sqrt{1-b^{2}u_R^{2}}}+\frac{\left(b^{2}u_S^{2}-2\right)}{\sqrt{1-b^{2}u_S^{2}}}\right] \nonumber \\
    &+\frac{3M}{\Delta r_s^{2}}\Bigg[\frac{b^{2}\left(-r_s^{2}u_R^{2}-2r_su_R+1\right)+2r_s^{2}}{b\sqrt{1-b^{2}u_R^{2}}}+\frac{b^{2}\left(-r_s^{2}u_S^{2}-2r_su_S+1\right)+2r_s^{2}}{b\sqrt{1-b^{2}u_S^{2}}} 
    +\frac{2r_s\arcsin(bu_R)}{\sqrt{1-b^{2}u_R^{2}}}+\frac{2r_s\arcsin(bu_S)}{\sqrt{1-b^{2}u_S^{2}}}\Bigg]\nonumber \\
    &-\frac{9mM}{\Delta r_s^{2}b}\Bigg[\frac{\left(2b^{2}+5r_s^{2}\right)\arcsin(bu_R)}{2b\left(1-b^{2}u_R^{2}\right)^{3/2}}+\frac{\left(2b^{2}+5r_s^{2}\right)\arcsin(bu_S)}{2b\left(1-b^{2}u_S^{2}\right)^{3/2}}\nonumber \\
    &-\frac{b^{2}\left(48u_R^{2}r_s+6u_R\right)+15u_Rr_s^{2}-32r_s}{6\left(1-b^{2}u_R^{2}\right)^{3/2}}-\frac{b^{2}\left(48u_S^{2}r_s+6u_S\right)+15u_Sr_s^{2}-32r_s}{6\left(1-b^{2}u_S^{2}\right)^{3/2}}\Bigg] \nonumber \\
    &+\frac{36\alpha m^{3}M}{L_*^{2}\Delta r_s^{2}b}\Bigg[\frac{\left(2b^{2}+5r_s^{2}\right)\arcsin(bu_R)}{2b\left(1-b^{2}u_R^{2}\right)^{3/2}}+\frac{\left(2b^{2}+5r_s^{2}\right)\arcsin(bu_S)}{2b\left(1-b^{2}u_S^{2}\right)^{3/2}} \nonumber \\ &-\frac{b^{2}\left(48u_R^{2}r_s+6u_R\right)+15u_Rr_s^{2}-32r_s}{6\left(1-b^{2}u_R^{2}\right)^{3/2}}-\frac{b^{2}\left(48u_S^{2}r_s+6u_S\right)+15u_Sr_s^{2}-32r_s}{6\left(1-b^{2}u_S^{2}\right)^{3/2}}\Bigg].
\end{align}
\end{widetext}

Using the inner product of the unit basis vector of the metric being considered, and the unit vector relative to the lensing object, we can find the important angles $\Psi_{S}$ and $\Psi_{R}$. The unit basis vector $e^i$, along the equatorial plane, is given by
\begin{equation} \label{e58}
    e^i=\left(\frac{dr}{dt},0,\frac{d\phi}{dt}\right)=\frac{d\phi}{dt}\left(\frac{dr}{d\phi},0,1\right)
\end{equation}
while the unit radial vector, which is along the radial direction from the lensing compact object is
\begin{equation} \label{e59}
    R^i=\left(\frac{1}{\sqrt{\gamma_{rr}}},0,0\right).
\end{equation}
Here, $\gamma_{rr}$ is the radial component of the optical metric which is defined in terms of metric coefficient as \cite{Ishihara_2016}
\begin{equation} \label{e60}
    \gamma_{ij}dx^i dx^j = \frac{1}{A(r)}\left(B(r)dr^2 + C(r) d\theta^2 + D(r,\theta) d\phi^2\right)
\end{equation}
if one sets $ds=0$. Focusing only along the equatorial plane, the inner product then suggest that
$$\cos \Psi\equiv\gamma_{ij}e^iR^j$$
\begin{equation} \label{e61}
    \cos \Psi=\sqrt{\gamma_{rr}}\frac{f(r)b}{r^2}\frac{dr}{d\phi}.
\end{equation}
Using $F(u)$ in Eq. \eqref{e51}, we find the most favorable function to find $\Psi$:
\begin{equation} \label{e62}
    \sin\Psi=\frac{f(r)^{1/2}b}{r}
\end{equation}
In the weak field limit, $\Psi$ for the source and receiver gives
\begin{widetext}
\begin{align} \label{e64}
    \Psi_R-\Psi_S & = \arcsin(bu_R)+\arcsin(bu_S)-\pi-bm\left[\frac{u_R^{2}}{\sqrt{1-b^{2}u_R^{2}}}+\frac{u_S^{2}}{\sqrt{1-b^{2}u_S^{2}}}\right] 
    -\frac{3bM}{\Delta r_s^{2}}\left[\frac{(u_Rr_s-1)^{2}}{\sqrt{1-b^{2}u_R^{2}}}+\frac{(u_Sr_s-1)^{2}}{\sqrt{1-b^{2}u_S^{2}}}\right] \nonumber\\
    &-\frac{2bM}{\Delta r_s^{3}}\left[\frac{(u_Rr_s-1)^{3}}{u_R\sqrt{1-b^{2}u_R^{2}}}+\frac{(u_Sr_s-1)^{3}}{u_S\sqrt{1-b^{2}u_S^{2}}}\right]-\frac{4\alpha bm^{3}}{L_*^{2}}\left[\frac{u_R^{2}}{\sqrt{1-b^{2}u_R^{2}}}+\frac{u_S^{2}}{\sqrt{1-b^{2}u_S^{2}}}\right] \nonumber\\
    &+\frac{3bmM}{\Delta r_s^{2}}\left[\frac{u_R\left(2b^{2}u_R^{2}-1\right)(u_Rr_s-1)^{2}}{\left(1-b^{2}u_R^{2}\right)^{3/2}}+\frac{u_S\left(2b^{2}u_S^{2}-1\right)(u_Sr_s-1)^{2}}{\left(1-b^{2}u_S^{2}\right)^{3/2}}\right] \nonumber\\
    &+\frac{2bmM}{\Delta r_s^{3}}\left[\frac{\left(2b^{2}u_R^{2}-1\right)(u_Rr_s-1)^{3}}{\left(1-b^{2}u_R^{2}\right)^{3/2}}+\frac{\left(2b^{2}u_S^{2}-1\right)(u_Sr_s-1)^{3}}{\left(1-b^{2}u_S^{2}\right)^{3/2}}\right] \nonumber\\
    &+\frac{12\alpha bm^{3}M}{L_*^{2}\Delta r_s^{2}}\left[\frac{u_R\left(2b^{2}u_R^{2}-1\right)(u_Rr_s-1)^{2}}{\left(1-b^{2}u_R^{2}\right)^{3/2}}+\frac{u_S\left(2b^{2}u_S^{2}-1\right)(u_Sr_s-1)^{2}}{\left(1-b^{2}u_S^{2}\right)^{3/2}}\right]\nonumber\\
    &+\frac{8\alpha bm^{3}M}{L_*^{2}\Delta r_s^{3}}\left[\frac{\left(2b^{2}u_R^{2}-1\right)(u_Rr_s-1)^{3}}{\left(1-b^{2}u_R^{2}\right)^{3/2}}+\frac{\left(2b^{2}u_S^{2}-1\right)(u_Sr_s-1)^{3}}{\left(1-b^{2}u_S^{2}\right)^{3/2}}\right].
\end{align}
After combining Eqs. \eqref{e57} and \eqref{e64}, we omit terms that depend on $1/\Delta r_s^3$, $1/L_*^2\Delta r_s^2$, $1/L_*^2\Delta r_s^3$ since their contribution is negligible and also noting that $M$ serves as additional effective mass to the black hole. The weak deflection angle, where $u_S$ and $u_R$ being finite distances is then
\begin{align} \label{e65}
    \hat{\alpha} & =\frac{2m}{b}\left(\sqrt{1-b^{2}u_{R}^{2}}+\sqrt{1-b^{2}u_{S}^{2}}\right)+\frac{6Mr_{s}^{2}}{b\Delta r_{s}^{2}}\left(\sqrt{1-b^{2}u_{R}^{2}}+\sqrt{1-b^{2}u_{S}^{2}}\right) \nonumber \\
    &-\frac{2bM}{\Delta r_{s}^3}\left[\frac{(r_{s}u_{R}-1)^{3}}{u_{R}\sqrt{1-b^{2}u_{R}^{2}}}+\frac{(r_{s}u_{S}-1)^{3}}{u_{S}\sqrt{1-b^{2}u_{S}^{2}}}\right]+\frac{8\alpha m^{3}}{bL_*^{2}}\left(\sqrt{1-b^{2}u_R^{2}}+\sqrt{1-b^{2}u_S^{2}}\right)
\end{align}
\end{widetext}

For the weak deflection angle of light in Eq. \eqref{e65}, we cannot impose $bu_R=0$ and $bu_S=0$ because the apparent divergence in the 3rd term. Since this implies that $u_R$ and $u_S$ must be finite, we can safely assume a far approximation such that $ u_{R} << 1 $ and $ u_{S} << 1 $. Thus we are left with
\begin{equation} \label{e66}
    \hat{\alpha} \approx \frac{4m}{b}+\frac{12Mr_s^{2}}{b\Delta r_s^{2}}+\frac{16\alpha m^{3}}{bL_*^{2}}+\frac{2bM}{u_R\Delta r_s^{3}}+\frac{2bM}{u_S\Delta r_s^{3}}.
\end{equation}
Eq. \eqref{e66} gives the expected Schwarzschild result for $\hat{\alpha}$ when both the dark matter and EUP correction are not present. Further, the result agrees in Ref. \cite{Kumaran:2019qqp} for EUP correction alone, and also agrees in Ref. \cite{Pantig_2020_weak} for dark matter contribution alone. Note that Eq. \eqref{e66} is the results from the second domain of the mass function in Eq. \eqref{e8}. Thus, the last two terms can vanish because of $1/\Delta r_s^3$ for the situation where $\Delta r_s>>1/u_S$ (or of $1/u_R$). In such a case, we have
\begin{equation} \label{e67} 
    \hat{\alpha} \approx \frac{4m}{b}+\frac{12Mr_s^{2}}{b\Delta r_s^{2}}+\frac{16\alpha m^{3}}{bL_*^{2}}.
\end{equation}

\begin{figure} [htpb] 
    \centering
    \includegraphics[width=\columnwidth]{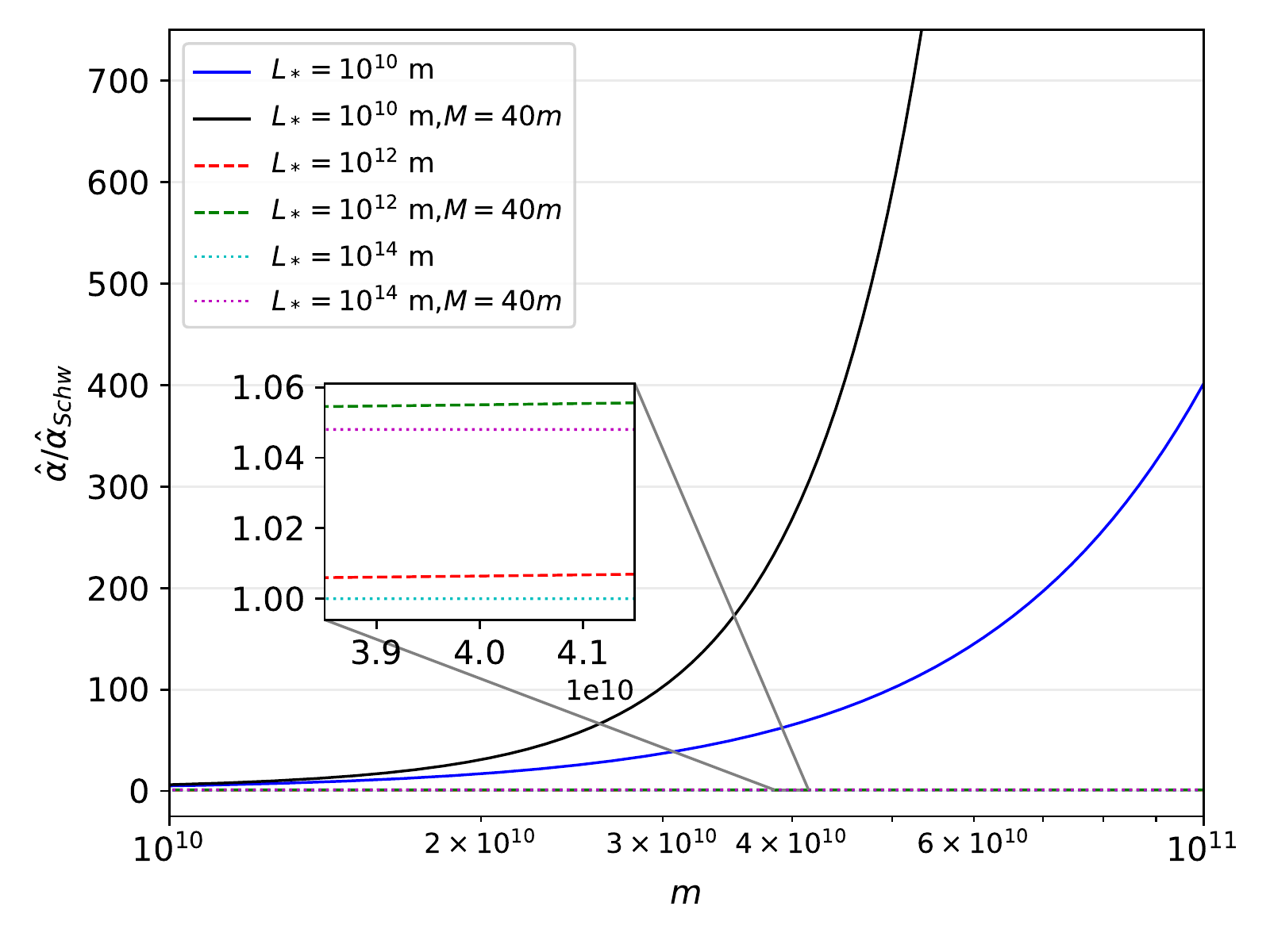}
    \caption{The normalized weak deflection angle vs. mass of the black hole $m$. Here, $\Delta r_s=100m$ and the impact parameter is set as $b=1000m$. The EUP parameter $\alpha$ is in unity.}
    \label{fig9}
\end{figure}

Fig. \ref{fig9} shows the plot of Eq. \eqref{e67} comparing EUP BH and EUP BH surrounded with dark matter. Here, the range of the black hole mass where we can see relevant differences is from $10^{10}$ m to $10^{11}$ m.
One can see that the immediate effect of the decreasing value of $L_*$ is to increase the value of the weak deflection angle for a given black hole mass $m$. The inset plot shows that $L_*=10^{14}$ m have negligible effects on the weak deflection angle, while a tiny deviation occurs when $L_*=10^{12}$ m. However, we can see considerable deviation that occurs when $L_*=10^{10}$ m. It can also be gleaned from the plot that the dark matter environment increases further the value of the weak deflection angle.

\begin{table*}
\scriptsize
\centering
\begin{tabular}{|l|p{1.6cm}|p{1.67cm}|p{1.67cm}|p{1.67cm}|p{1.6cm}|p{1.6cm}|p{1.6cm}|}
\hline
\multicolumn{8}{|c|}{Weak deflection angle [$\mu$as]}                                                                                                               \\ 
\hline
\multicolumn{1}{|c|}{\multirow{2}{*}{}} & \multirow{2}{*}{Schw. Term} & \multicolumn{3}{c|}{DM contribution} & \multicolumn{3}{c|}{EUP contribution}  \\  
\cline{3-8} \\[-1em]
\multicolumn{1}{|c|}{}                  &                             & $L_*=10^{10}$m & $L_*=10^{12}$m & $L_*=10^{14}$m & $L_*=10^{10}$m & $L_*=10^{12}$m & $L_*=10^{14}$m   \\ 
\hline 
Sgr. A*                                  & $825.05922$                           & $2.47083\text{e}(-14)$          & $3.61893\text{e}(-15)$          & $3.61776\text{e}(-15)$          & $1331.12967$          & $0.13311$          & $1.33113\text{e}(-5)$            \\
M87                                     & $825.05922$                           & $110.59855$          & $1.13432\text{e}(-6)$          & $1.82163\text{e}(-10)$          & $6.47927\text{e}(8)$          & $6.47927\text{e}(4)$          & $6.47927$            \\
UGC 7232                                & $825.05922$                           & $3.44460\text{e}(-15)$           & $3.43859\text{e}(-15)$          & $3.43859\text{e}(-15)$         & $0.71992$          & $7.19919\text{e}(-5)$          & $7.19919\text{e}(-9)$            \\
\hline
\end{tabular}
\caption{Numerical values of the weak deflection angle obtained from Eq. \eqref{e67}. Here, the impact parameter used is $b=1000m$. Other parameters are the same as that in Table \ref{tab:table1}.}
\label{tab:table2}
\end{table*}

Using Eq. \eqref{e67}, let us now see the effects of dark matter and EUP contribution to the weak deflection angle. Consider first our Sun, where $M_{\odot}=1477$ m, and radius $R_{\odot}=6.96\text{x}10^9$ m, which is also the impact parameter for convenience. A calculation will show that the Schwarzschild term is $\sim 1.75 \mu$as, as expected. However, if $L_*=10^{10}$ m is used, the EUP contribution is only $\sim 1.53\text{x}10^{-13}\mu$as, which is very small to be detected by the current sensitivity of astronomical instruments available. Furthermore, the deviation due to dark matter effect is smaller: $\sim 9.92\text{x}10^{-14}\mu$as. In Table \ref{tab:table2}, we listed the numerical values of the weak deflection angle of the black hole at the center of different galaxies. For the Schwarzschild term, the values are the same. This indicates that the only way the weak deflection angle changes is when the impact parameter changes. Note also that this value is an estimate since the black hole is not rotating. For the DM contribution, we see that the deviations are not fixed since the second term in Eq. \eqref{e67} depends on $r_s$, which in turn also depends on $L_*$. It indicates that the weak deflection angle is more sensitive to deviation than the shadow radius. For the EUP contribution, the deviation is fully dependent on the black hole's mass. For example, using the black hole in M87, the dark matter effect is hopeless to be detected at $L_*=10^{14}$ m. Nevertheless, if this is the true fundamental length scale, there is hope for the EUP effect to be detected if the equipment is sensitive enough at $6.48\mu$as. The Event Horizon Telescope can achieve an angular resolution of $10\mu$as to map the stellar neighborhood near the black hole at the center of our galaxy. In mapping our whole galaxy, the ESA's GAIA mission can provide a sensitivity of $20\mu$as to $7\mu$as depending on the stellar magnitudes \cite{Liu_2017}. Finally, we remark that for M87, $L_*=10^{10}$ m can indeed be ruled out because of an abnormal increase of angle in the EUP contribution.

Using Eq. \eqref{e67}, we can form an estimate to $\Delta r_s$ in a similar manner to the estimate made using the shadow radius. For considerable changes to the weak deflection angle to occur, the dark matter thickness requirement should be
\begin{equation} \label{e69}
    \Delta r_s=\frac{2\sqrt{3mM}\left(1+\frac{4\alpha m^{2}}{L_*^{2}}\right)}{\sqrt{1-\frac{4\alpha m^{2}}{L_*^{2}}}}.
\end{equation}
Note that such requirement is improved by a factor of $2(1+4\alpha m^2/L_*^2)$. However, the restriction that $L_* > 2m$ still occurs and if $L_*>>m$, then $\Delta r_s = 2\sqrt{3mM}$, a result that is in agreement with Ref.  \cite{Pantig_2020_weak}. Without an EUP correction, the values of $\Delta r_s$ for Sgr A* and M87 are $1.84$x$10^{13}$ m and $1.49$x$10^{15}$ m respectively. With the EUP correction in Eq. \eqref{e69}, these estimated values increases to $1.64$x$10^{14}$ m and $5.72$x$10^{15}$ m respectively. We used the estimated values of $m$ and $M$ presented in the previous section. According to this result, notable deviation due to dark matter and EUP correction is not evident to the weak deflection angle since $\Delta r_s$ is not satisfied to these galaxies. Indeed, $\Delta r_s$ is in many orders of magnitude smaller compared to the characteristic size of the galaxy, or the estimated size of the dark matter halo that envelops them.

\section{Conclusion} \label{sec5}
In this article, we have studied the effect of dark matter on a quantum black hole whose quantum fluctuations spewed over very large distances due to the fundamental length scale $L_*$, which at present has an unknown value. A study in Ref. \cite{Lu_2019} constrained the values $L_*$ using the gravitational lensing observables indicating that $L_*\sim10^{10}$ m for the SMBH at Sgr. A* and $L_*\sim10^{13}$ m for M87 galaxy. Of course, there must be only one value for the large fundamental length scale (the same way we only have one value for the Planck length), and one of these constraints must be ruled out. It is also possible, in theory, that $L_*$ might be the Hubble's length which gives interesting consequences to the masses comparable to galactic scales \cite{Mureika_2019}.

This study has shown how the effects of $L_*$ vary inversely to the mass of the compact object being considered. For example, if we probe the SMBH at M87, the use of $L_*=10^{10}$ m gives large deviations that can be ruled out by current data and observation. On the other hand, if $L_*=10^{14}$ m is used, considerable deviations are present that might be detectable by future observatories. If it happens to be that $L_*$ is the Hubble's length, then seeking its effect on the SMBH at the M87 galaxy might be negligible. Moreover, since we have a quantum model of a black hole based on its mass, we surrounded it with dark matter and explored its possible consequences to the black hole geometry. The result of the analysis revealed that time-like particles are more sensitive to deviation despite low dark matter density (see Figs. \ref{nfig4} and \ref{fig4}). Null particles, which give manifestation to the black hole shadow, require more dark matter mass to show considerable deviation. Comparing Figs. \ref{fig5} and \ref{fig6}, there is a difference between the EUP and dark matter effects. For a range in the black hole mass where the effect of $L_*$ begins to manifest, increasing $L_*$ decreases the value of both the photon sphere and shadow radii. The role played by the dark matter, however, is different since it decreases the value of the photon sphere radius while increasing that of the shadow radius. Such behavior for dark matter can be accounted for being an astrophysical environment, where light must travel through it. Using Eq. \eqref{e50}, we formed an estimate of the dark matter thickness ($\Delta r_s$) required to show considerable deviation to the shadow radius. With $L_*$ present, the required $\Delta r_s$ increases but still not enough to be satisfied to a certain galaxy. Hence, although there is a possibility of detecting the effects $L_*$, detecting dark matter effects using the shadow of a black hole at the center of a galaxy remains hopeless.

We have also shown the interplay between $L_*$ and dark matter mass to the weak deflection angle. We found out that increasing the value of $L_*$ decreases the amount of deviation, which is in contrast to the dark matter effect (see Fig. \ref{fig9}). Such a difference can be seen clearly in Table \ref{tab:table2}. Using the M87 galaxy, improvements in angular resolution of astronomical observatories can potentially confirm EUP effects due to $L_*=10^{14}$ m, but still hopelessly detect dark matter effects. Using $L_*=10^{10}$ m to M87 galaxy gives a ridiculous amount of deviation due to dark matter and EUP effects, hence, such a value for $L_*$ can indeed be ruled out. The derived expression for the weak deflection angle also allowed one to estimate the dark matter thickness $\Delta r_s$ as a condition to observe a noticeable effect to the weak deflection angle. Although the estimate has been improved through the inclusion of the EUP parameters $\alpha$ and $L_*$, and better compared to shadow radius, realistic values of $m$ and $M$ indicate that the change in the weak deflection angle due to the EUP and dark matter is still not satisfied.

Finally, we note some shortcomings of the model. The dark matter mass considered in this study is static, uniform, and ignores the gravitational pull of the black hole because what matters in the analysis is the mass of the dark matter. Future investigation may include a realistic scenario of non-uniform dark matter distribution, and how its dynamics near the black hole might affect the black hole geometry. Moreover, future research direction may also include the EUP correction to a rotating black hole. The EUP alone may affect the galactic dynamics, but it is also interesting to examine the corresponding effect of dark matter on it. When the value of the large fundamental length scale is established, reexamination of such a EUP model is also needed. For completeness, the Generalized Uncertainty Principle (GUP) parameter can also be included in the metric and explore the possible consequences.

\section{Acknowledgement} \label{sec6}
We are very grateful to the anonymous referee, whose insightful comments and suggestions contribute to the betterment of this paper.

\section{Appendix A}
In this appendix, we show that the EUP black hole surrounded by dark matter described by the metric function given in Eq. \eqref{e11} is an exact solution to the Einstein field equation. With the pure Schwarzschild metric, $T_{\mu\nu}=0$ and it immediately satisfies the Einstein field equation. The metric in this study, however, incorporates dark matter as an astrophysical environment and a modification in the black hole's mass using the extended uncertainty principle. Thus, one cannot expect that $T_{\mu\nu} \ne 0$ since the spacetime is no longer asymptotically flat due to the dark matter shell. To begin, consider a non-zero stress-energy tensor
\begin{equation}
    T_{\mu\nu}=\delta_{\mu}^{\alpha}\delta_{\nu}^{\beta}G_{\alpha\beta}.
\end{equation}
which is needed to satisfy the Einstein field equation
\begin{equation}
    G_{\mu\nu}-8\pi\delta_{\mu}^{\alpha}\delta_{\nu}^{\beta}G_{\alpha\beta}=0.
\end{equation}
Here, the Kronecker delta is related to the orthogonal bases, so we can write
\begin{equation}
    e_{a}^{\mu}e_{\alpha}^{a}T_{\mu\nu}=e_{b}^{\beta}e_{\nu}^{b}G_{\alpha\beta} \rightarrow e_{a}^{\mu}e_{b}^{\nu}T_{\mu\nu}=e_{a}^{\alpha}e_{b}^{\beta}G_{\alpha\beta}
\end{equation}
after the indices are switched. If we define $T_{ab}=e_{a}^{\mu}e_{b}^{\nu}T_{\mu\nu}$, the components of the stress-energy tensor are then
\begin{equation}
    T_{ab}=e_{a}^{\alpha}e_{b}^{\beta}G_{\alpha\beta}=(\rho, p_{r}, p_{\theta}, p_{\phi}).
\end{equation}
The components of the Einstein tensor are then the following:
\begin{align}
    &G_{tt}=\frac{2\mathcal{M}'(r)}{r^{2}}f(r), \nonumber \\
    &G_{rr}=-\frac{2\mathcal{M}'(r)}{r^{2}}f(r)^{-1}, \nonumber \\
    &G_{\theta \theta}=-\mathcal{M}''(r)r, \nonumber \\
    &G_{\phi \phi}=\mathcal{M}''(r)r\sin^2 \theta.
\end{align}
As mentioned earlier about the orthogonal bases, the convenient choice is the following:
\begin{align} 
    &e_{t}^{\alpha}=\frac{1}{\sqrt{f(r)}}\left(1,0,0,0\right), \nonumber \\
    &e_{r}^{\alpha}=\sqrt{f(r)}\left(0,1,0,0\right), \nonumber \\
    &e_{\theta}^{\alpha}=\frac{1}{r}\left(0,0,1,0\right), \nonumber \\
    &e_{\phi}^{\alpha}=\frac{1}{r\sin \theta}\left(0,0,0,1\right).
\end{align}
Using the two previous equations, the components of $T_{ab}$ can now be determined:
\begin{align}
    &\rho=-p_{r}=\frac{2\mathcal{M}'(r)}{8\pi r^2}, \nonumber \\
    &p_{\theta}=p_{\phi}=p_{r}-\frac{2\mathcal{M}'(r)+\mathcal{M}''(r)r}{8\pi r^2}.
\end{align}
Since the EUP mass function does not depend on $r$, $\mathcal{M}'(r)$ and $\mathcal{M}''(r)$ depends only on the dark matter model in the study. As a final remark, we don't know the full mechanism of dark matter as well as the gravitons inside the event horizon to develop the appropriate Lagrangian in an underlying field theory. Such a limitation has also been the same with the published work in \cite{Jusufi_2019,Hou_2018,Xu_2018} where the metric incorporating the dark matter profile came from empirical observation. Although the determination of the Lagrangian and the calculation of the action may provide potential future work, this is not beyond the scope of this paper, however.

\bibliography{references}

\end{document}